%% file: 2025.tex
\documentclass[epjc3,twocolumn]{svjour3}

\usepackage{currfile} 
\usepackage{subcaption}
\usepackage{stfloats}
\usepackage{cuted}
\usepackage{placeins}
\usepackage[colorlinks=true,citecolor=blue,filecolor=blue,linkcolor=blue,urlcolor=blue,pdftex]{hyperref}
\usepackage{orcidlink} 
\graphicspath{{./figs/}}
\usepackage{caption}
\usepackage{multirow} 
\usepackage{lineno}
\usepackage{amsmath, amssymb}
\usepackage{graphicx}

\journalname{Noname}

\title{{Model-independent searches of new physics in DARWIN with deep learning}}

\author{J.~Aalbers\orcidlink{0000-0003-0030-0030}\thanksref{addr:groningen}
\and
K.~Abe\orcidlink{0009-0000-9620-788X}\thanksref{addr:tokyo}
\and
M.~Adrover\thanksref{addr:zurich}
\and
S.~Ahmed Maouloud\orcidlink{0000-0002-0844-4576}\thanksref{addr:paris}
\and
L.~Althueser\orcidlink{0000-0002-5468-4298}\thanksref{addr:munster}
\and
D.~W.~P.~Amaral\orcidlink{0000-0002-1414-932X}\thanksref{addr:rice}
\and
B.~Andrieu\orcidlink{0009-0002-6485-4163}\thanksref{addr:paris}
\and
E.~Angelino\orcidlink{0000-0002-6695-4355}\thanksref{addr:torino,addr:lngs}
\and
D.~Ant\'on Martin\thanksref{addr:chicago}
\and
B.~Antunovic\thanksref{addr:belgrade,addr:banjaluka}
\and
E.~Aprile\orcidlink{0000-0001-6595-7098}\thanksref{addr:columbia}
\and
M.~Babicz\thanksref{addr:zurich}
\and
D.~Bajpai\thanksref{addr:UniversityofAlabama}
\and
M.~Balzer\thanksref{addr:kitipe}
\and
E.~Barberio\thanksref{addr:melbourne}
\and
L.~Baudis\orcidlink{0000-0003-4710-1768}\thanksref{addr:zurich}
\and
M.~Bazyk\orcidlink{0009-0000-7986-153X}\thanksref{addr:subatech,addr:melbourne}
\and
N.~F.~Bell\thanksref{addr:melbourne}
\and
L.~Bellagamba\orcidlink{0000-0001-7098-9393}\thanksref{addr:bologna}
\and
R.~Biondi\orcidlink{0000-0002-6622-8740}\thanksref{addr:mpik}
\and
Y.~Biondi\thanksref{addr:kit}
\and
A.~Bismark\orcidlink{0000-0002-0574-4303}\thanksref{addr:zurich}
\and
C.~Boehm\thanksref{addr:sydney}
\and
K.~Boese\orcidlink{0009-0007-0662-0920}\thanksref{addr:mpik}
\and
R.~Braun\thanksref{addr:munster}
\and
A.~Breskin\thanksref{addr:wis}
\and
S.~Brommer\thanksref{addr:kitetp}
\and
A.~Brown\orcidlink{0000-0002-1623-8086}\thanksref{addr:freiburg,addr:Sheffield}
\and
G.~Bruni\orcidlink{0000-0001-5667-7748}\thanksref{addr:bologna}
\and
R.~Budnik\orcidlink{0000-0002-1963-9408}\thanksref{addr:wis}
\and
C.~Cai\thanksref{addr:tsinghua}
\and
C.~Capelli\orcidlink{0000-0003-3330-621X}\thanksref{addr:zurich}
\and
A.~Chauvin\thanksref{addr:heidelberg}
\and
A.~P.~Cimental~Chavez\orcidlink{0009-0004-9605-5985}\thanksref{addr:zurich}
\and
A.~P.~Colijn\orcidlink{0000-0002-3118-5197}\thanksref{addr:nikhef}
\and
J.~Conrad\orcidlink{0000-0001-9984-4411}\thanksref{addr:stockholm}
\and
J.~J.~Cuenca-Garc\'ia\orcidlink{0000-0002-3869-7398}\thanksref{addr:zurich}
\and
V.~D'Andrea\orcidlink{0000-0003-2037-4133}\thanksref{addr:lngs,addr:roma}
\and
L.~C.Daniel~Garcia\orcidlink{0009-0000-5813-9118}\thanksref{addr:paris}
\and
M.~P.~Decowski\orcidlink{0000-0002-1577-6229}\thanksref{addr:nikhef}
\and
A.~Deisting\orcidlink{0000-0001-5372-9944}\thanksref{addr:mainz}
\and
C.~Di~Donato\orcidlink{0009-0005-9268-6402}\thanksref{addr:laquila}
\and
P.~Di~Gangi\orcidlink{0000-0003-4982-3748}\thanksref{addr:bologna}
\and
S.~Diglio\orcidlink{0000-0002-9340-0534}\thanksref{addr:subatech}
\and
M.~Doerenkamp\thanksref{addr:heidelberg}
\and
G.~Drexlin\thanksref{addr:kitetp}
\and
K.~Eitel\orcidlink{0000-0001-5900-0599}\thanksref{addr:kit}
\and
A.~Elykov\orcidlink{0000-0002-2693-232X}\thanksref{addr:kit}
\and
R.~Engel\thanksref{addr:kit}
\and
A.~D.~Ferella\orcidlink{0000-0002-6006-9160}\thanksref{addr:laquila,addr:lngs}
\and
C.~Ferrari\orcidlink{0000-0002-0838-2328}\thanksref{addr:lngs}
\and
H.~Fischer\orcidlink{0000-0002-9342-7665}\thanksref{addr:freiburg}
\and
T.~Flehmke\orcidlink{0009-0002-7944-2671}\thanksref{addr:stockholm}
\and
M.~Flierman\orcidlink{0000-0002-3785-7871}\thanksref{addr:nikhef}
\and
K.~Fujikawa\thanksref{addr:nagoya}
\and
W.~Fulgione\orcidlink{0000-0002-2388-3809}\thanksref{addr:torino,addr:lngs}
\and
C.~Fuselli\orcidlink{0000-0002-7517-8618}\thanksref{addr:nikhef}
\and
P.~Gaemers\orcidlink{0009-0003-1108-1619}\thanksref{addr:nikhef}
\and
R.~Gaior\orcidlink{0009-0005-2488-5856}\thanksref{addr:paris}
\and
M.~Galloway\orcidlink{0000-0002-8323-9564}\thanksref{addr:zurich}
\and
F.~Gao\orcidlink{0000-0003-1376-677X}\thanksref{addr:tsinghua}
\and
N.~Garroum\thanksref{addr:paris}
\and
R.~Giacomobono\orcidlink{0000-0001-6162-1319}\thanksref{addr:napels}
\and
F.~Girard\thanksref{addr:paris}
\and
R.~Glade-Beucke\orcidlink{0009-0006-5455-2232}\thanksref{addr:freiburg}
\and
F.~Gl\"uck\thanksref{addr:kit}
\and
L.~Grandi\orcidlink{0000-0003-0771-7568}\thanksref{addr:chicago}
\and
J.~Grigat\orcidlink{0009-0005-4775-0196}\thanksref{addr:freiburg}
\and
R.~Gr\"o{\ss}le\thanksref{addr:kit}
\and
H.~Guan\orcidlink{0009-0006-5049-0812}\thanksref{addr:purdue}
\and
M.~Guida\orcidlink{0000-0001-5126-0337}\thanksref{addr:mpik}
\and
P.~Gyorgy\orcidlink{0009-0005-7616-5762 }\thanksref{addr:mainz}
\and
R.~Hammann\orcidlink{0000-0001-6149-9413}\thanksref{addr:mpik}
\and
V.~Hannen\thanksref{addr:munster}
\and
S.~Hansmann-Menzemer\thanksref{addr:heidelberg}
\and
N.~Hargittai\thanksref{addr:wis}
\and
A.~Higuera\orcidlink{0000-0001-9310-2994}\thanksref{addr:rice}
\and
C.~Hils\thanksref{addr:mainz}
\and
K.~Hiraoka\thanksref{addr:nagoya}
\and
L.~Hoetzsch\orcidlink{0000-0003-2572-477X}\thanksref{addr:mpik}
\and
N.~F.~Hood\orcidlink{0000-0003-2507-7656}\thanksref{addr:ucsd}
\and
M.~Iacovacci\orcidlink{0000-0002-3102-4721}\thanksref{addr:napels}
\and
Y.~Itow\orcidlink{0000-0002-8198-1968}\thanksref{addr:nagoya}
\and
J.~Jakob\thanksref{addr:munster}
\and
R.~S.~James\thanksref{addr:melbourne,addr:UniversityCollegeLondonUCL}
\and
F.~Joerg\orcidlink{0000-0003-1719-3294}\thanksref{addr:mpik,addr:zurich}
\and
F.~Kahlert\orcidlink{0009-0000-1500-9794}\thanksref{addr:purdue}
\and
Y.~Kaminaga\orcidlink{0009-0006-5424-2867}\thanksref{addr:tokyo}
\and
M.~Kara\orcidlink{0009-0004-5080-9446}\thanksref{addr:kit}
\and
P.~Kavrigin\orcidlink{0009-0000-1339-2419}\thanksref{addr:wis}
\and
S.~Kazama\orcidlink{0000-0002-6976-3693}\thanksref{addr:nagoya}
\and
M.~Keller\thanksref{addr:heidelberg}
\and
P.~Kharbanda\thanksref{addr:nikhef}
\and
B.~Kilminster\thanksref{addr:zurich}
\and
M.~Kleifges\thanksref{addr:kitipe}
\and
M.~Klute\thanksref{addr:kitetp}
\and
M.~Kobayashi\thanksref{addr:nagoya}
\and
D.~Koke\thanksref{addr:munster}
\and
A.~Kopec\orcidlink{0000-0001-6548-0963}\thanksref{addr:bucknell}
\and
B.~von~Krosigk\thanksref{addr:heidelbergki}
\and
F.~Kuger\orcidlink{0000-0001-9475-3916}\thanksref{addr:freiburg}
\and
L.~LaCascio\thanksref{addr:kitetp}
\and
H.~Landsman\orcidlink{0000-0002-7570-5238}\thanksref{addr:wis}
\and
R.~F.~Lang\orcidlink{0000-0001-7594-2746}\thanksref{addr:purdue}
\and
L.~Levinson\orcidlink{0000-0003-4679-0485}\thanksref{addr:wis}
\and
I.~Li\orcidlink{0000-0001-6655-3685}\thanksref{addr:rice}
\and
A.~Li\thanksref{addr:ucsd}
\and
S.~Li\orcidlink{0000-0003-0379-1111}\thanksref{addr:westlake}
\and
S.~Liang\orcidlink{0000-0003-0116-654X}\thanksref{addr:rice}
\and
Z.~Liang\thanksref{addr:shenzhen}
\and
Y.~-T.~Lin\thanksref{addr:mpik}
\and
S.~Lindemann\orcidlink{0000-0002-4501-7231}\thanksref{addr:freiburg}
\and
M.~Lindner\orcidlink{0000-0002-3704-6016}\thanksref{addr:mpik}
\and
K.~Liu\thanksref{addr:tsinghua}
\and
J.~Loizeau\thanksref{addr:subatech}
\and
F.~Lombardi\orcidlink{0000-0003-0229-4391}\thanksref{addr:mainz}
\and
J.~Long\orcidlink{0000-0002-5617-7337}\thanksref{addr:chicago}
\and
J.~A.~M.~Lopes\orcidlink{0000-0002-6366-2963}\thanksref{addr:coimbra,addr:coimbrapoli}
\and
G.~M.~Lucchetti\orcidlink{0000-0003-4622-036X}\thanksref{addr:bologna}
\and
T.~Luce\orcidlink{8561-4854-7251-585X}\thanksref{addr:freiburg}
\and
Y.~Ma\orcidlink{0000-0002-5227-675X}\thanksref{addr:ucsd}
\and
C.~Macolino\orcidlink{0000-0003-2517-6574}\thanksref{addr:laquila,addr:lngs}
\and
J.~Mahlstedt\orcidlink{0000-0002-8514-2037}\thanksref{addr:stockholm}
\and
B.~Maier\thanksref{addr:kitetp,addr:ImperialCollegeLondon}
\and
A.~Mancuso\orcidlink{0009-0002-2018-6095}\thanksref{addr:bologna}
\and
L.~Manenti\orcidlink{0000-0001-7590-0175}\thanksref{addr:sydney}
\and
F.~Marignetti\orcidlink{0000-0001-8776-4561}\thanksref{addr:napels}
\and
K.~Martens\orcidlink{0000-0002-5049-3339}\thanksref{addr:tokyo}
\and
J.~Masbou\orcidlink{0000-0001-8089-8639}\thanksref{addr:subatech}
\and
E.~Masson\orcidlink{0000-0002-5628-8926}\thanksref{addr:paris}
\and
S.~Mastroianni\orcidlink{0000-0002-9467-0851}\thanksref{addr:napels}
\and
A.~Melchiorre\orcidlink{0009-0006-0615-0204}\thanksref{addr:laquila}
\and
J.~Men\'endez\orcidlink{0000-0002-1355-4147}\thanksref{addr:barcelona}
\and
M.~Messina\orcidlink{0000-0002-6475-7649}\thanksref{addr:lngs}
\and
B.~Milosovic\thanksref{addr:belgrade}
\and
S.~Milutinovic\thanksref{addr:belgrade}
\and
K.~Miuchi\orcidlink{0000-0002-1546-7370}\thanksref{addr:kobe}
\and
R.~Miyata\thanksref{addr:nagoya}
\and
A.~Molinario\orcidlink{0000-0002-5379-7290}\thanksref{addr:torino}
\and
C.~M.~B.~Monteiro\thanksref{addr:coimbra}
\and
K.~Mor\aa\orcidlink{0000-0002-2011-1889}\thanksref{addr:columbia}
\and
S.~Moriyama\orcidlink{0000-0001-7630-2839}\thanksref{addr:tokyo}
\and
E.~Morteau\thanksref{addr:subatech}
\and
Y.~Mosbacher\thanksref{addr:wis}
\and
J.~M\"uller\orcidlink{0009-0007-4572-6146}\thanksref{addr:freiburg}
\and
M.~Murra\orcidlink{0009-0008-2608-4472}\thanksref{addr:columbia}
\and
J.~L.~Newstead\thanksref{addr:melbourne}
\and
K.~Ni\orcidlink{0000-0003-2566-0091}\thanksref{addr:ucsd}
\and
C.~O'Hare\thanksref{addr:sydney}
\and
U.~Oberlack\orcidlink{0000-0001-8160-5498}\thanksref{addr:mainz}
\and
M.~Obradovic\thanksref{addr:belgrade}
\and
I.~Ostrowskiy\thanksref{addr:UniversityofAlabama}
\and
S.~Ouahada\thanksref{addr:zurich}
\and
B.~Paetsch\orcidlink{0000-0002-5025-3976}\thanksref{addr:wis}
\and
Y.~Pan\orcidlink{0000-0002-0812-9007}\thanksref{addr:paris}
\and
M.~Pandurovic\thanksref{addr:belgrade}
\and
Q.~Pellegrini\orcidlink{0009-0002-8692-6367}\thanksref{addr:paris}
\and
R.~Peres\orcidlink{0000-0001-5243-2268}\thanksref{addr:zurich}
\and
F.~Piastra\orcidlink{0000-0001-8848-5089}\thanksref{addr:zurich}
\and
J.~Pienaar\orcidlink{0000-0001-5830-5454}\thanksref{addr:chicago,addr:wis}
\and
M.~Pierre\orcidlink{0000-0002-9714-4929}\thanksref{addr:nikhef}
\and
G.~Plante\orcidlink{0000-0003-4381-674X}\thanksref{addr:columbia}
\and
T.~R.~Pollmann\orcidlink{0000-0002-1249-6213}\thanksref{addr:nikhef}
\and
L.~Principe\orcidlink{0000-0002-8752-7694}\thanksref{addr:subatech,addr:melbourne}
\and
J.~Qi\orcidlink{0000-0003-0078-0417}\thanksref{addr:ucsd}
\and
K.~Qiao\thanksref{addr:nikhef}
\and
J.~Qin\orcidlink{0000-0001-8228-8949}\thanksref{addr:rice}
\and
M.~Rajado\orcidlink{0000-0002-7663-2915}\thanksref{addr:zurich}
\and
D.~Ram\'irez~Garc\'ia\orcidlink{0000-0002-5896-2697}\thanksref{addr:zurich}
\and
A.~Ravindran\orcidlink{0009-0004-6891-3663}\thanksref{addr:subatech,addr:melbourne}
\and
A.~Razeto\thanksref{addr:lngs}
\and
L.~Sanchez\thanksref{addr:rice}
\and
P.~Sanchez-Lucas\thanksref{addr:zurich,addr:grenada}
\and
G.~Sartorelli\orcidlink{0000-0003-1910-5948}\thanksref{addr:bologna}
\and
A.~Scaffidi\thanksref{addr:sissa,emailA}
\and
J.~Schreiner\thanksref{addr:mpik}
\and
P.~Schulte\orcidlink{0009-0008-9029-3092}\thanksref{addr:munster}
\and
H.~Schulze Ei{\ss}ing\orcidlink{0009-0005-9760-4234}\thanksref{addr:munster}
\and
M.~Schumann\orcidlink{0000-0002-5036-1256}\thanksref{addr:freiburg}
\and
A.~Schwenck\thanksref{addr:kit}
\and
L.~Scotto~Lavina\orcidlink{0000-0002-3483-8800}\thanksref{addr:paris}
\and
M.~Selvi\orcidlink{0000-0003-0243-0840}\thanksref{addr:bologna}
\and
F.~Semeria\orcidlink{0000-0002-4328-6454}\thanksref{addr:bologna}
\and
P.~Shagin\thanksref{addr:mainz}
\and
S.~Sharma\thanksref{addr:heidelberg}
\and
W.~Shen\thanksref{addr:heidelberg}
\and
S.~Y.~Shi\orcidlink{0000-0002-2445-6681}\thanksref{addr:columbia}
\and
T.~Shimada\thanksref{addr:nagoya}
\and
H.~Simgen\orcidlink{0000-0003-3074-0395}\thanksref{addr:mpik}
\and
R.~Singh\orcidlink{0000-0001-9564-7795}\thanksref{addr:purdue}
\and
M.~Solmaz\thanksref{addr:heidelbergki,addr:kitetp}
\and
O.~Stanley\thanksref{addr:melbourne,addr:subatech}
\and
M.~Steidl\thanksref{addr:kit}
\and
A.~Stevens\orcidlink{0009-0002-2329-0509}\thanksref{addr:freiburg}
\and
A.~Takeda\orcidlink{0009-0003-6003-072X}\thanksref{addr:tokyo}
\and
P.-L.~Tan\orcidlink{0000-0002-5743-2520}\thanksref{addr:stockholm}
\and
D.~Thers\orcidlink{0000-0002-9052-9703}\thanksref{addr:subatech}
\and
T.~Th\"ummler\thanksref{addr:kit}
\and
F.~T\"onnies\orcidlink{0000-0002-2287-5815}\thanksref{addr:freiburg}
\and
F.~Toschi\thanksref{addr:kit}
\and
G.~Trinchero\thanksref{addr:torino}
\and
R.~Trotta\orcidlink{0000-0002-3415-0707}\thanksref{addr:sissa,addr:ImperialCollegeLondon,emailR}
\and
C.~D.~Tunnell\orcidlink{0000-0001-8158-7795}\thanksref{addr:rice}
\and
P.~Urquijo\thanksref{addr:melbourne}
\and
M.~Utoyama\thanksref{addr:nagoya}
\and
K.~Valerius\orcidlink{0000-0001-7964-974X}\thanksref{addr:kit}
\and
S.~Vecchi\orcidlink{0000-0002-4311-3166}\thanksref{addr:ferrara}
\and
S.~Vetter\orcidlink{0009-0001-2961-5274}\thanksref{addr:kit}
\and
G.~Volta\orcidlink{0000-0001-7351-1459}\thanksref{addr:mpik}
\and
D.~Vorkapic\thanksref{addr:belgrade}
\and
W.~Wang\thanksref{addr:UniversityofAlabama}
\and
K.~M.~Weerman\thanksref{addr:nikhef}
\and
C.~Weinheimer\orcidlink{0000-0002-4083-9068}\thanksref{addr:munster}
\and
M.~Weiss\orcidlink{0009-0005-3996-3474}\thanksref{addr:wis}
\and
D.~Wenz\thanksref{addr:munster}
\and
M.~Wilson\thanksref{addr:kit}
\and
C.~Wittweg\orcidlink{0000-0001-8494-740X}\thanksref{addr:zurich}
\and
J.~Wolf\thanksref{addr:kitetp}
\and
V.~H.~S.~Wu\orcidlink{0000-0002-8111-1532}\thanksref{addr:kit}
\and
S.~W\"ustling\thanksref{addr:kitipe}
\and
M.~Wurm\thanksref{addr:mainz}
\and
Y.~Xing\orcidlink{0000-0002-1866-5188}\thanksref{addr:melbourne}
\and
D.~Xu\orcidlink{0000-0001-7361-9195}\thanksref{addr:columbia}
\and
Z.~Xu\orcidlink{0000-0002-6720-3094}\thanksref{addr:columbia}
\and
M.~Yamashita\orcidlink{0000-0001-9811-1929}\thanksref{addr:tokyo}
\and
L.~Yang\orcidlink{0000-0001-5272-050X}\thanksref{addr:ucsd}
\and
J.~Ye\orcidlink{0000-0002-6127-2582}\thanksref{addr:shenzhen}
\and
L.~Yuan\orcidlink{0000-0003-0024-8017}\thanksref{addr:chicago}
\and
G.~Zavattini\orcidlink{0000-0002-6089-7185}\thanksref{addr:ferrara}
\and
M.~Zhong\thanksref{addr:ucsd}
\and
K.~Zuber\orcidlink{0000-0001-8689-4495}\thanksref{addr:dresden}
(XLZD Collaboration\thanksref{email1}). }
\newcommand{\banjaluka}{University of Banja Luka, 78000 Banja Luka, Bosnia and Herzegovina}
\newcommand{\barcelona}{Department of Quantum Physics and Astrophysics and Institute of Cosmos Sciences, University of Barcelona, 08028 Barcelona, Spain}
\newcommand{\belgrade}{Vinca Institute of Nuclear Science, University of Belgrade, Mihajla Petrovica Alasa 12-14. Belgrade, Serbia}
\newcommand{\bern}{Albert Einstein Center for Fundamental Physics, Institute for Theoretical Physics, University of Bern, Sidlerstrasse 5, 3012 Bern, Switzerland}

\newcommand{\bologna}{Department of Physics and Astronomy, University of Bologna and INFN-Bologna, 40126 Bologna, Italy}

\newcommand{\bucknell}{Department of Physics \& Astronomy, Bucknell University, Lewisburg, PA, USA}

\newcommand{\chicago}{Department of Physics, Enrico Fermi Institute \& Kavli Institute for Cosmological Physics, University of Chicago, Chicago, IL 60637, USA}
\newcommand{\coimbra}{LIBPhys, Department of Physics, University of Coimbra, 3004-516 Coimbra, Portugal}
\newcommand{\coimbrapoli}{Coimbra Polytechnic - ISEC, 3030-199 Coimbra, Portugal}
\newcommand{\columbia}{Physics Department, Columbia University, New York, NY 10027, USA}
\newcommand{\darmstadt}{Department of Physics, Technische Universita\"at Darmstadt, 64289 Darmstadt, Germany}
\newcommand{\dresden}{Technische Universit\"at Dresden, 01069 Dresden, Germany}

\newcommand{\ferrara}{INFN-Ferrara and Dip. di Fisica e Scienze della Terra, Universit\`a di Ferrara, 44122 Ferrara, Italy}
\newcommand{\freiburg}{Physikalisches Institut, Universit\"at Freiburg, 79104 Freiburg, Germany}
\newcommand{\freiburgBrown}{Physikalisches Institut, Universit\"at Freiburg, 79104 Freiburg, Germany (Now at Sheffield)}
\newcommand{\grenada}{University of Grenada}
\newcommand{\groningen}{Nikhef and the University of Groningen, Van Swinderen Institute, 9747AG Groningen, Netherlands}
\newcommand{\heidelberg}{Physikalisches Institut, Universit\"at Heidelberg, Heidelberg, Germany}
\newcommand{\heidelbergki}{Kirchhoff-Institut f\"ur Physik, Universit\"at Heidelberg, Heidelberg, Germany}
\newcommand{\ImperialCollegeLondon}{Physics Department, Imperial College London Blackett Laboratory, London SW7 2AZ, UK}

\newcommand{\kit}{Institute for Astroparticle Physics, Karlsruhe Institute of Technology, 76021 Karlsruhe, Germany}
\newcommand{\kitetp}{Institute of Experimental Particle Physics, Karlsruhe Institute of Technology, 76021 Karlsruhe, Germany}
\newcommand{\kitipe}{Institute for Data Processing and Electronics, Karlsruhe Institute of Technology, 76021 Karlsruhe, Germany}
\newcommand{\kobe}{Department of Physics, Kobe University, Kobe, Hyogo 657-8501, Japan}

\newcommand{\laquila}{Department of Physics and Chemistry, University of L'Aquila, 67100 L'Aquila, Italy}

\newcommand{\lngs}{INFN-Laboratori Nazionali del Gran Sasso and Gran Sasso Science Institute, 67100 L'Aquila, Italy}
\newcommand{\mainz}{Institut f\"ur Physik \& Exzellenzcluster PRISMA$^{+}$, Johannes Gutenberg-Universit\"at Mainz, 55099 Mainz, Germany}
\newcommand{\melbourne}{ARC Centre of Excellence for Dark Matter Particle Physics, School of Physics, The University of Melbourne, VIC 3010, Australia}
\newcommand{\mpik}{Max-Planck-Institut f\"ur Kernphysik, 69117 Heidelberg, Germany}
\newcommand{\munster}{Institute for Nuclear Physics, University of M\"unster, 48149 M\"unster, Germany}
\newcommand{\nagoya}{Kobayashi-Maskawa Institute for the Origin of Particles and the Universe, and Institute for Space-Earth Environmental Research, Nagoya University, Furo-cho, Chikusa-ku, Nagoya, Aichi 464-8602, Japan}
\newcommand{\napels}{Department of Physics ``Ettore Pancini'', University of Napoli and INFN-Napoli, 80126 Napoli, Italy}

\newcommand{\nikhef}{Nikhef and the University of Amsterdam, Science Park, 1098XG Amsterdam, Netherlands}
\newcommand{\paris}{LPNHE, Sorbonne Universit\'{e}, CNRS/IN2P3, 75005 Paris, France}

\newcommand{\purdue}{Department of Physics and Astronomy, Purdue University, West Lafayette, IN 47907, USA}
\newcommand{\rice}{Department of Physics and Astronomy, Rice University, Houston, TX 77005, USA}
\newcommand{\roma}{INFN-Roma Tre, 00146 Roma, Italy}

\newcommand{\shenzhen}{School of Science and Engineering, The Chinese University of Hong Kong, Shenzhen, Guangdong, 518172, P.R. China}
\newcommand{\sissa}{Theoretical and Scientific Data Science, Scuola Internazionale Superiore di Studi Avanzati (SISSA), 34136 Trieste, Italy}

\newcommand{\stockholm}{Oskar Klein Centre, Department of Physics, Stockholm University, AlbaNova, Stockholm SE-10691, Sweden}
\newcommand{\subatech}{SUBATECH, IMT Atlantique, CNRS/IN2P3,  Nantes Universit\'e, Nantes 44307, France}
\newcommand{\sydney}{School of Physics, The University of Sydney, Camperdown, Sydney, NSW 2006, Australia}
\newcommand{\tokyo}{Kamioka Observatory, Institute for Cosmic Ray Research, and Kavli Institute for the Physics and Mathematics of the Universe (WPI), University of Tokyo, Higashi-Mozumi, Kamioka, Hida, Gifu 506-1205, Japan}
\newcommand{\torino}{INAF-Astrophysical Observatory of Torino, Department of Physics, University  of  Torino and  INFN-Torino,  10125  Torino,  Italy}
\newcommand{\tsinghua}{Department of Physics \& Center for High Energy Physics, Tsinghua University, Beijing 100084, P.R. China}
\newcommand{\ucsd}{Department of Physics, University of California San Diego, La Jolla, CA 92093, USA}
\newcommand{\UniversityCollegeLondonUCL}{Department of Physics and Astronomy, University College London (UCL), London WC1E 6BT, UK}
\newcommand{\UniversityofAlabama}{Department of Physics \& Astronomy, University of Alabama, Tuscaloosa, AL 34587-0324, USA}

\newcommand{\UniversityofSheffield}{Department of Physics and Astronomy, University of Sheffield, Sheffield S3 7RH, UK}

\newcommand{\westlake}{Department of Physics, School of Science, Westlake University, Hangzhou 310030, P.R. China}
\newcommand{\wis}{Department of Particle Physics and Astrophysics, Weizmann Institute of Science, Rehovot 7610001, Israel}
\newcommand{\zurich}{Physik-Institut, University of Z\"urich, 8057  Z\"urich, Switzerland}
\authorrunning{XLZD Collaboration}
\thankstext{addr:banjaluka}{Also at \banjaluka}\hypertarget{addr:banjaluka}{}
\thankstext{addr:roma}{Also at \roma}\hypertarget{addr:roma}{}
\thankstext{addr:coimbrapoli}{Also at \coimbrapoli}\hypertarget{addr:coimbrapoli}{}
\thankstext{addr:grenada}{Also at \grenada}\hypertarget{addr:grenada}{}
\thankstext{addr:Sheffield}{Now at \UniversityofSheffield}\hypertarget{addr:UniversityofSheffield}{}

\thankstext{email1}{\texttt{xlzd@xlzd.org}}\hypertarget{email1}{}
\thankstext{emailA}{\texttt{ascaffid@sissa.it}}\hypertarget{emailA}{}
\thankstext{emailR}{\texttt{rtrotta@sissa.it}}\hypertarget{emailR}{}

\institute{\hypertarget{addr:groningen}{\groningen}\label{addr:groningen}
\and
\hypertarget{addr:tokyo}{\tokyo}\label{addr:tokyo}
\and
\hypertarget{addr:zurich}{\zurich}\label{addr:zurich}
\and
\hypertarget{addr:paris}{\paris}\label{addr:paris}
\and
\hypertarget{addr:munster}{\munster}\label{addr:munster}
\and
\hypertarget{addr:rice}{\rice}\label{addr:rice}
\and
\hypertarget{addr:torino}{\torino}\label{addr:torino}
\and
\hypertarget{addr:lngs}{\lngs}\label{addr:lngs}
\and
\hypertarget{addr:chicago}{\chicago}\label{addr:chicago}
\and
\hypertarget{addr:belgrade}{\belgrade}\label{addr:belgrade}
\and
\hypertarget{addr:columbia}{\columbia}\label{addr:columbia}
\and
\hypertarget{addr:UniversityofAlabama}{\UniversityofAlabama}\label{addr:UniversityofAlabama}
\and
\hypertarget{addr:kitipe}{\kitipe}\label{addr:kitipe}
\and
\hypertarget{addr:melbourne}{\melbourne}\label{addr:melbourne}
\and
\hypertarget{addr:subatech}{\subatech}\label{addr:subatech}
\and
\hypertarget{addr:bologna}{\bologna}\label{addr:bologna}
\and
\hypertarget{addr:mpik}{\mpik}\label{addr:mpik}
\and
\hypertarget{addr:kit}{\kit}\label{addr:kit}
\and
\hypertarget{addr:sydney}{\sydney}\label{addr:sydney}
\and
\hypertarget{addr:wis}{\wis}\label{addr:wis}
\and
\hypertarget{addr:kitetp}{\kitetp}\label{addr:kitetp}
\and
\hypertarget{addr:freiburg}{\freiburg}\label{addr:freiburg}
\and
\hypertarget{addr:freiburgBrown}{\freiburgBrown}\label{addr:freiburgBrown}
\and
\hypertarget{addr:UniversityofSheffield}{\UniversityofSheffield}\label{addr:UniversityofSheffield}
\and
\hypertarget{addr:tsinghua}{\tsinghua}\label{addr:tsinghua}
\and
\hypertarget{addr:heidelberg}{\heidelberg}\label{addr:heidelberg}
\and
\hypertarget{addr:nikhef}{\nikhef}\label{addr:nikhef}
\and
\hypertarget{addr:stockholm}{\stockholm}\label{addr:stockholm}
\and
\hypertarget{addr:mainz}{\mainz}\label{addr:mainz}
\and
\hypertarget{addr:laquila}{\laquila}\label{addr:laquila}
\and
\hypertarget{addr:nagoya}{\nagoya}\label{addr:nagoya}
\and
\hypertarget{addr:napels}{\napels}\label{addr:napels}
\and
\hypertarget{addr:purdue}{\purdue}\label{addr:purdue}
\and
\hypertarget{addr:bern}{\bern}\label{addr:bern}
\and
\hypertarget{addr:ucsd}{\ucsd}\label{addr:ucsd}
\and
\hypertarget{addr:UniversityCollegeLondonUCL}{\UniversityCollegeLondonUCL}\label{addr:UniversityCollegeLondonUCL}
\and
\hypertarget{addr:bucknell}{\bucknell}\label{addr:bucknell}
\and
\hypertarget{addr:heidelbergki}{\heidelbergki}\label{addr:heidelbergki}
\and
\hypertarget{addr:westlake}{\westlake}\label{addr:westlake}
\and
\hypertarget{addr:shenzhen}{\shenzhen}\label{addr:shenzhen}
\and
\hypertarget{addr:coimbra}{\coimbra}\label{addr:coimbra}
\and
\hypertarget{addr:ImperialCollegeLondon}{\ImperialCollegeLondon}\label{addr:ImperialCollegeLondon}
\and
\hypertarget{addr:barcelona}{\barcelona}\label{addr:barcelona}
\and
\hypertarget{addr:kobe}{\kobe}\label{addr:kobe}
\and
\hypertarget{addr:sissa}{\sissa}\label{addr:sissa}
\and
\hypertarget{addr:darmstadt}{\darmstadt}\label{addr:darmstadt}
\and
\hypertarget{addr:ferrara}{\ferrara}\label{addr:ferrara}
\and
\hypertarget{addr:dresden}{\dresden}\label{addr:dresden}
}
\usepackage{xcolor}

\begin{document}

\onecolumn 
\maketitle
\twocolumn
\begin{abstract}
We present a deep learning pipeline to perform a model-independent, likelihood-free search for anomalous (i.e., non-background) events in the proposed next-generation multi-ton scale liquid xenon-based direct detection experiment, DARWIN. We train an anomaly detector comprising a variational autoencoder (VAE) and a classifier on high-dimensional simulated detector response data and construct a 1D anomaly score to reject the background-only hypothesis in the presence of an excess of non-background-like events. We use simulated validation data to determine the power of the method to reject the background-only hypothesis in the presence of a WIMP dark matter signal, without any model-dependent assumption about the nature of the signal. We show that our neural networks learn relevant features of the events from low-level, high-dimensional detector outputs, avoiding lossy and computationally expensive compression into lower-dimensional observables.  {Our approach is complementary to the usual likelihood-based analysis, in that it reduces the reliance on many of the corrections and cuts that are traditionally part of the analysis chain, with the potential of achieving higher accuracy and significant reduction of analysis time. We envisage the methodology presented in this work augmenting or complementing likelihood-based and other data-driven methods currently utilized in the DARWIN (and in the future, XLZD) analysis pipeline. }
\end{abstract}

\section{Introduction} \label{sec:intro}

A promising method for investigations of the ever-elusive dark matter sector involves seeking excess nuclear recoils in subterranean detectors, a strategy known as direct detection (DD)~\cite{Goodman:1984dc}. Over the years, a number of xenon   (XENONnT~\cite{XENON:2023cxc}, LUX-ZEPLIN (LZ)~\cite{Aalbers_2023LZ}, PandaX\cite{PandaX-4T:2021bab}) and argon (DEAP-3600~\cite{Lai:2023qub}, DarkSide-20k~\cite{Aalseth:2017fik}, ArDM~\cite{Calvo:2016hve}) ton-scale experiments have steadily enhanced the sensitivity to physics beyond the standard model (BSM), and this effort is expected to continue, with plans for a next-generation dark matter and neutrino observatory. While earlier designs for a `dark matter WIMP search with liquid xenon' observatory (DARWIN)~\cite{Aalbers:2016jon,Aalbers_2023} aimed at an active liquid xenon target mass of 40 tons, the recently formed XLZD Collaboration proposes an even more ambitious target mass of 60–80 tons \cite{xlzdDB}. While the design of the XLZD experiment is being developed, this paper focuses on DARWIN, a well-defined proposal for a large-scale observatory using a xenon dual-phase time projection chamber (TPC) to study phenomena requiring low-background conditions. DARWIN aims to be sensitive to weakly interacting massive particle (WIMP) dark matter as well as neutrinoless double beta decay, axion-like particles, and any other BSM particles that would manifest through significant interaction with a xenon target. The aim of this work is to introduce a signal model-agnostic, deep learning-based analysis pipeline,  {offering a complementary and alternative approach} to the standard likelihood-based analysis chain in such a detector. The benefits of this approach are that it enables a fuller exploitation of the detector readout data, without the information loss potentially incurred in using only hand-crafted summary statistics (such as cS1 and cS2, the corrected prompt primary scintillation and
secondary electroluminescence of ionized electrons signals, respectively), and that it can include in the pipeline any physics effect that can be faithfully simulated, including systematics.  

Machine learning (ML) has emerged as a powerful tool within the physics community, and its relevance to DM phenomenology has been growing rapidly \cite{Carleo_2019,Zhang:2019ryt,Lucie-Smith:2019hdl,Bernardini:2019bmd,Todarello:2023qrr}. 

Unsupervised machine learning has been increasingly employed in collider physics to identify anomalies in data, as demonstrated in several recent studies~\cite{Farina:2018fyg,Dery:2018dqr,Collins:2018epr,Otten:2019hhl,Blance_2019,Blance_2021,Heimel:2018mkt,Kuusela:2011aa,knapp2020adversarially,Andreassen:2020nkr,Nachman:2020lpy,Collins:2019jip}, with early example applications on simulated events of CMS and ATLAS already in Refs.~\cite{cerri2019variational,vanBeekveld:2020txa}, as well as Ref.~\cite{Khosa:2020qrz}, where an ``anomaly awareness" algorithm is proposed. ML techniques were also applied to DD experiments for a variety of tasks, ranging from signal classification to fast likelihood evaluation~\cite{Coarasa:2022zak,us_sup,Akerib2022,DarkSide-50:2023fcw,XENONCollaboration:2023dar,Lopez-Fogliani:2024gzj}. Ref.~\cite{us_sup} utilizes a semi-unsupervised deep neural network comprising a pretrained convolutional neural network (CNN) and a VAE in order to detect the presence of excess nuclear recoils above the expected background in DD experiments.

The established approach to the detection of a new physics signal in DD experiment with dual dual-phase target is a likelihood-based test with an assumed asymptotic distribution~\cite{Aalbers_2023}, with the likelihood a function of the so-called ``corrected'' S1 and S2 signals (cS1 and cS2, respectively). By using neural networks that are trained on high-dimensional representations of detector events, we show in this paper that it is possible to infer the relevant properties (energy distribution, type of recoil) from detector-level readouts, without the approximation and loss of information incurred in the usual cS1, cS2 compression.  {This opens the door to the possibility of an end-to-end inference approach that is fully simulation-based, including all necessary corrections and cuts that are traditionally done in the analysis and inference chain, a process which takes up a significant fraction of analysis time in current-generation detectors. This approach relies however, on the availability of accurate and faithful simulations: real detectors and backgrounds are usually more complex and/or feature unexpected characteristics that deviate from simulations. Data-driven calibration and adversarial training techniques can help mitigate such systematic differences, improving robustness against these biases -- something we plan to explore in future works.} 

 {Subject to the above caveat, the aim of this paper is to demonstrate the capability of a deep learning pipeline to detect the presence of an `anomalous' signal above a known (from simulations) background in DARWIN, without explicit modeling of the likelihood nor of the physics underlying the anomaly (i.e., without assuming a specific dark matter model). In this sense, our analysis is model-independent, that is, agnostic to any specific new physics model. We achieve this by training an anomaly detector on event-by-event simulated detector response quanta using the DARWIN simulation pipeline, and by constructing an anomaly score designed to maximize the sensitivity to rejecting the background-only hypothesis. The choice of DARWIN as a case study is motivated by the availability of sufficiently mature and complete detector simulations, which is not yet the case for XLZD. Of course, the general approach is applicable to future detectors, once their design and simulation pipeline are settled. Application of this approach to existing detectors would require refinement to account for rare and/or unforeseen backgrounds or detector effects that may not be simulated correctly. Since this paper focuses on demonstrating the overall methodology, we leave exploration of such issues to future investigations.}  

This paper is structured as follows. In Sec.~\ref{sec:DARWIN} we briefly introduce the design of the DARWIN detector, we describe the data structure used to train the model, as well as the simulations that were employed to this end. In Sec.~\ref{sec:methodology} we explain the aim of the analysis, the methodology employed and its novelty. We also present the detailed simulation pipeline adopted for the study, the split between training and validation sets and the training procedure. In Sec.~\ref{sec:analysis}, we validate our approach by determining the sensitivity of DARWIN to rejecting the background-only null-hypothesis in the presence of a simulated injection of a WIMP signal. We then conclude in Sec.~\ref{sec:conclusion}.

\section{Experiment design and data simulation}
\label{sec:DARWIN}
\subsection{The DARWIN detector design}

 {DARWIN is conceived as a multi–ton, dual–phase liquid xenon time–projection
chamber (TPC) designed to push DD sensitivity to the verge of
the astrophysical neutrino floor~\cite{chepel2013liquid}.  The reference design holds
$\sim50\,$t of xenon, with about $40\,$t active, inside a $2.6\;\text{m}\times2.6\;\text{m}$
cylindrical TPC; prompt VUV scintillation (S1) and proportional
electroluminescence (S2) are captured by matched top and bottom arrays of
ultra--low--background photomultipliers (PMTs) or {silicon photomultiplier (SiPM)} tiles, providing
sub--keV thresholds and event--by--event electron vs nuclear recoil
discrimination.   {The large homogeneous target, excellent
self--shielding and simultaneous light--and--charge readout make large TPC chambers versatile platforms for dark matter, neutrino and rare decay
physics~\cite{Aalbers:2016jon}.} }

 {The TPC design is suspended in a double--walled low--radio\-activity cryostat and
immersed in an instrumented water tank that serves both as a passive
$\gamma / n$ shield and an active Cherenkov muon veto.
A uniform drift field of the order of 0.5 kV cm$^{-1}$ is generated inside the TPC, enabling electrons to traverse the full 2.6 m height.  This long-drift capability- as well as cryogenics,
purification, and DAQ concepts,  {has been} validated in the {Xenoscope}
vertical demonstrator and related optical simulation
test--stands~\cite{Peres_2023}, as well as a second large scale demonstrator called PANCAKE \cite{pancake}.   }

 {In 2024 the DARWIN, \textsc{LZ} and {XENONnT}
collaborations unified their efforts in the next-generation {XLZD}
programme~\cite{xlzdDB}, which scales the dual--phase concept to $60$--$80\,$t of active
xenon while retaining the core detector architecture.
DARWIN’s hardware prototypes and simulation tools remain the
principal testbeds for XLZD component development and the
waveform-level analysis showcased here. Consequently, the study performed in this paper adopts the original 40~t DARWIN geometry when generating simulated
S1/S2 events, with the ML methodology and data analysis pipeline having direct application to any future XLZD-type detector.}

\subsection{Generation of Simulated Events}
\begin{figure*}[t]
    \centering
    \begin{subfigure}[t]{0.5\textwidth}
        \centering
        \includegraphics[width=\textwidth]{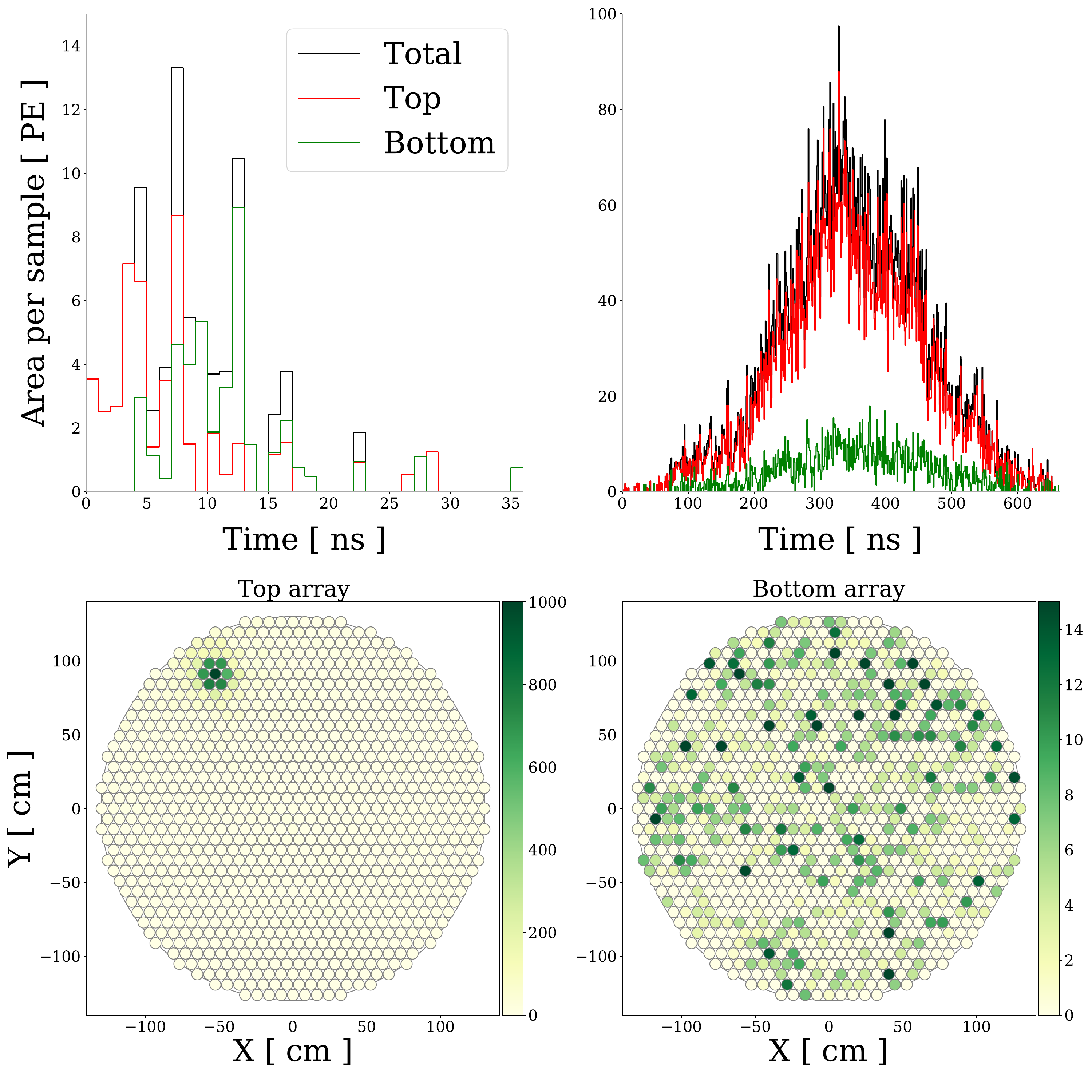}
        \caption{ER event}
        \label{fig:ER_event}
    \end{subfigure}%
    \begin{subfigure}[t]{0.5\textwidth}
        \centering
        \includegraphics[width=\textwidth]{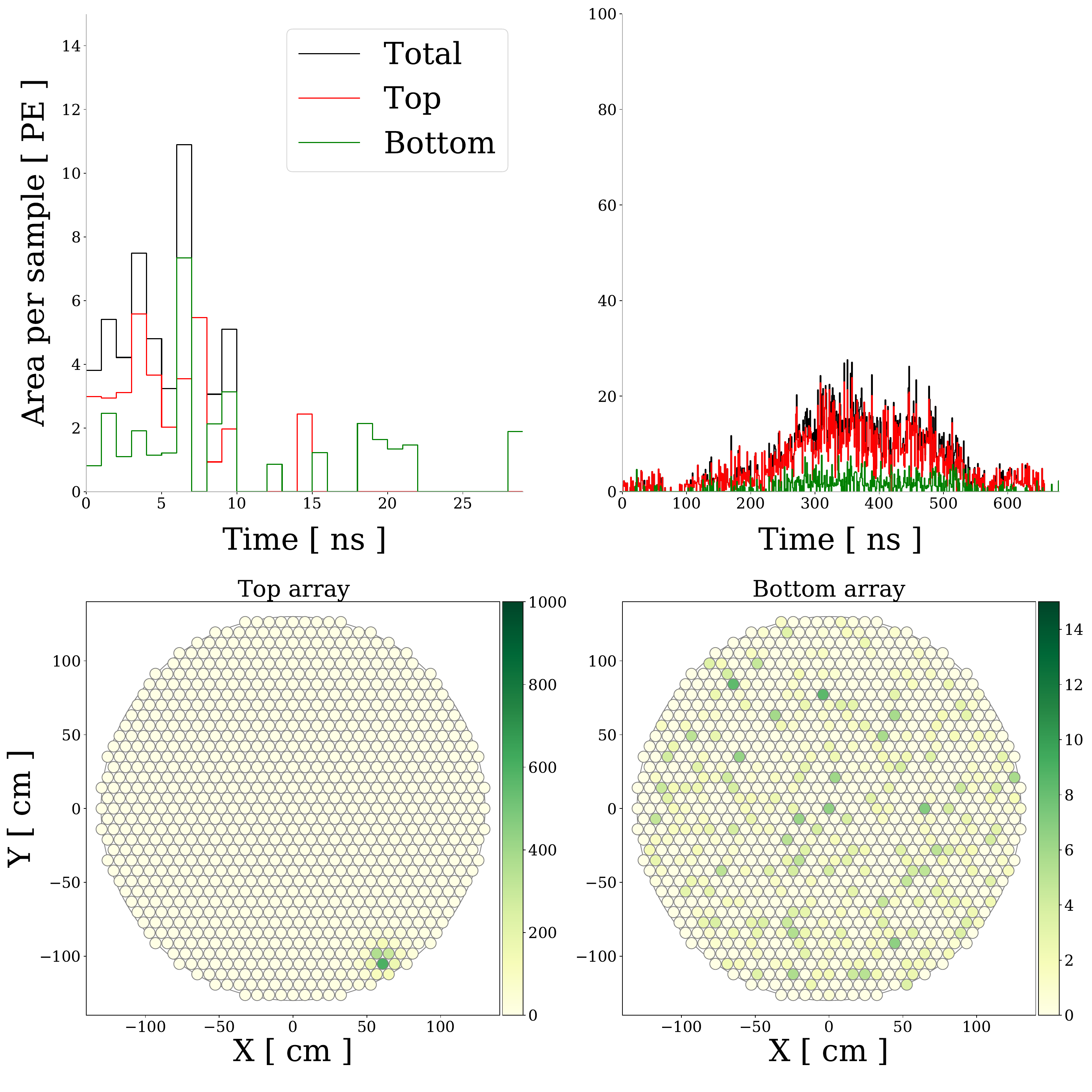}
        \caption{NR event}
        \label{fig:NR_event}
    \end{subfigure}
    \caption{Example of simulated detector observables of an electron recoil (ER) (a) and nuclear recoil (NR) (b) event in DARWIN. \textbf{Top}: Number of S1 (left sub-panel) and S2 (right sub-panel) photoelectrons (PE) as a function of time after initial S1 triggering. Red (green) denotes observation in the top (bottom) PMT array. The black curves are the total S1 + S2 and are used for training the neural networks. \textbf{Bottom}: Top and bottom S2 PMT deposit spatial pattern. The color bar indicates the PMT hit count. These data are used to train the neural networks.}
    \label{fig:data_example}
\end{figure*}

 {Our simulation-based pipeline is reliant on the quality of the simulations adopted. For this reason, we use state-of-the-art simulations tailored to the DARWIN design.} We use the Geant4 transport code \cite{GEANT4:2002zbu} within the DARWIN-Geant4 framework \cite{DARWIN:2023uje} to handle the tracking of particles within a rendering of the detector geometry. The Noble Element Simulation Technique (\texttt{NEST}) \texttt{v2.3.12}~\cite{nest2018} handles the microphysics of how particles interact with the active xenon volume. \texttt{NEST} provides a robust and well-established framework that simulates the atomic and nuclear physics involved in energy deposition and the corresponding response of the detector, and generates the light and charge yields for each type of interaction within the detector. These simulated light and charge yields are compared and calibrated against previous xenon experiments, see Ref.~\cite{Aalbers_2023} for details. Full signal propagation and observable readout within the TPC volume that produced the simulated waveforms and PMT hit-patterns were produced by custom-written detector simulation code based on the \texttt{Tray}~\cite{Abbasi_2022} architecture.

 {Any WIMP search relies on distinguishing between background events and the WIMP-induced signal. We therefore need our deep learning pipeline to learn to characterize the background distribution.} The majority of background at DARWIN will be electron recoil (ER) events originating from various terrestrial and cosmogenic sources, while nuclear recoil (NR) backgrounds remain in the form of irreducible cosmogenic neutrinos and sub-dominant radiogenic neutrons~\cite{Schumann_2015,DARWIN:2023uje}, which must be included as part of the background simulation. WIMPs of mass $\mathcal{O}(>1)$ GeV deposit their energy into the detector via NR events. 

 {We describe the background simulations used in this study in \ref{sec:bkg}, and give here only a concise summary. For each type of background (ER and NR), events with uniformly distributed recoil energies were simulated in the range 1-100 keV.  The simulations include detector response effects (including electron-ion recombination, electron drift, and photon-collection efficiency), which transform the raw energy deposition from the initial particle interaction into the observable signals in the detector.}\footnote{Another form of background observed by XENONnT was anomalous events emanating from radioactivity in the TPC walls, referred to as 'surface' events. We did not simulate such event realizations in this study, but work is being conducted to implement them with unsupervised veto models. }

  {For our analysis, we follow the approach taken in Ref.~\cite{us_sup}, and adopt as description of the TPC data the total  S1 + S2 waveforms (i.e, signal as a function of time, summed over all individual PMTs), as well as the top and bottom S2 PMT hit pattern readout\footnote{We note that not including the S1 top and bottom PMT hit-patterns decreases sensitivity to so-called `$\gamma$-X' and `neutron-X' events as observed by XENON100 \cite{weber2013,Kessler:2014jya}. Since we do not include such background events in this study, this has no bearing on our results here. Future developments will add both `$\gamma$-X' and `neutron-X' events to the background, and S1 PMT patterns to the input data.}. We use the total waveforms (as opposed to the PMT-specific waveform) in order to reduce the dimensionality and complexity of the data vector provided to the neural networks. To exploit the detector readout data in even more fundamental form, one should adopt a model capable of learning a representation of the PMT responses from the entire PMT array in the temporal domain \cite{XENON:2022vye,Kessler:2014jya} -- something the method in this work is unable to scale to. Modern developments in Transformer or graph neural network architectures could potentially be used for handling time-domain individual PMT readouts \cite{hewitt2024deep,Jiang:2024wph,Farrell:2024aah}. In order to meet this challenge however, we plan to utilize the Rotary Masked Autoencoder of Ref.~\cite{2025arXiv250520535Z}.} 
 
 In Fig.~\ref{fig:data_example} we show an example of the data used to train the neural networks. Events are simulated in a fiducial detector volume (FV) of 31.5 t, chosen to optimize the detection of rare NR while minimizing ER background interference towards the boundaries of the bulk xenon, as well as other factors \cite{Aalbers_2023}. The simulations are realized with a drift field of 200.0 V/cm, registering events when at least 4 photons are detected within a 200-nanosecond window (referred to as a `4-fold coincidence', or N4T200). We do not utilize spatial reconstruction to provide a further fiducialization cut. Work is being done in this direction at XENON, see for example Ref.~\cite{vetter2024}.

\section{Methodology}
\label{sec:methodology}

 {In this section, we first provide an overview of the objective of this study, followed by a concise description of the analysis methodology, which highlights the novelty of the approach. The architectural details as well as hyperparameters of the VAE and classifier used in this study are detailed in \ref{sec:arch}. }

\subsection{Simulation-based anomaly detection}
\label{sec:anom_detec_intro}

 {The objective of this study is to demonstrate the potential of a deep learning pipeline to detect a WIMP-like signal above known simulated backgrounds in a semi-supervised fashion. This is complementary to the traditional likelihood-based method, as it offers several potential advantages: first, our approach makes fuller use of the information contained in the PMT readout data, thus avoiding the information loss that compression into summary statistics (such as cS1/cS2) inevitably incurs; secondly, it can incorporate in the pipeline any effect that can be faithfully simulated in the mock data. This means that the impact of nuisance parameters can be accounted for by simply including their sampling within the generation of training data. Finally, our approach does not rely on approximations to the likelihood, nor to a model-specific form of the WIMP-signal, therefore being more general and model-agnostic. }

 {Our aim is to train a suitable neural network to identify anomalous signals -- i.e., any event that can be distinguished statistically from the simulated ER and NR background distribution.} This involves the computation of an `anomaly score', $TS$, obtained from the combined loss distribution and classification output of a neural anomaly detector. The anomaly score is used to ascertain whether a collection of observed events ${\bf X}_n = \{ \textbf{x}_1, \textbf{x}_2,\dots, \textbf{x}_n\}, $ deviates from the background-only distribution. The null hypothesis, which we denote $\mathcal{H}_0$, is that the events $\textbf{X}_n$ are drawn from a distribution where no signal is present, i.e., compatible with the expected background. 
 \begin{figure}[!t]
    \centering
    \includegraphics[width=0.5\textwidth]{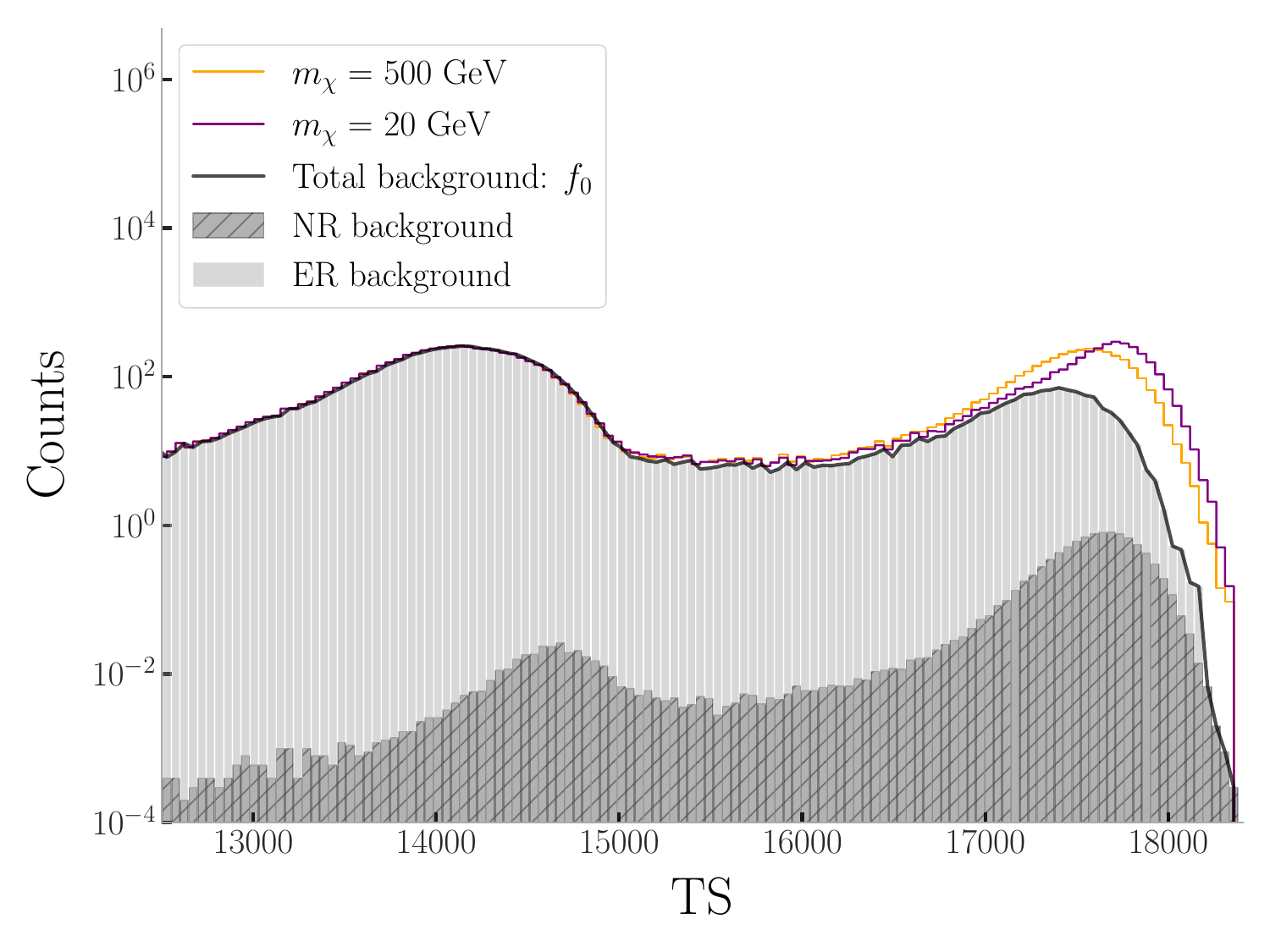}
    \caption{Distribution of the anomaly score $TS$ from a pseudo-dataset used in this study. The stacked gray bars represent the  $TS$ distribution for the ER (light gray) and NR (dark gray) background.   The colored lines are the distributions in $TS$ after the injection of signal components for 20 and 500 GeV WIMPs, with a scattering cross-section of $\sigma_\chi = 10^{-46}$ cm$^2$ (a large value chosen for clarity of illustration). The binning is illustrative, as our sensitivity analysis is unbinned. The solid black line is the total background pdf $f_0$.}
    \label{fig:p_dataset}
\end{figure}

 {The anomaly detector consists of two parts: a supervised binary classifier and a VAE. The classifier learns from training data to distinguish ER from NR events, whilst the VAE is trained solely on ER events\footnote{An alternative approach (see, for example, the work of the LUX-ZEPLIN (LZ) collaboration in
 Ref.~\cite{Arthurs2024}) can be to train a VAE on a representative sample of \textit{all} event classes (comprising both ER and NR) as well as calibration data. This allows potentially anomalous events to be identified in the latent space of the autoencoder. Representation learning of this type can be useful in that it is one-sample and 'data-driven', at the expense of sensitivity to an explicit null-hypothesis. }. After training, validation data (i.e., that the network has not been trained on) is given to the network, and its $TS$ distribution obtained: events that deviated from background-like properties will manifest in the 1D space of the $TS$ distribution as an excess over the background-only distribution. A simple 1D statistical test is then employed to reject the background-only hypothesis. }

\subsection{Definition and distribution of the anomaly score}
\label{sec:anom_score}
The anomaly score, $TS$, is defined as the weighted linear combination of the reconstruction loss from the VAE, or `ELBO' (see Eqn.~\ref{eqn:recon_loss}), and the classifier's binary cross-entropy, $H_B$, so that larger values correspond to deviations from the null hypothesis:
\begin{align}
  TS &=  (-\text{ELBO}) + RH_B \nonumber\\
  \nonumber \\
     &= D_\text{KL}(q(\mathbf{z} | \mathbf{x}_\text{in}) || p(\mathbf{z}))  - \mathbb{E}_{q(\mathbf{z}|\mathbf{x}_{\text{in}})}[\log p_{\mathbf{x}_{\text{in}}}(\mathbf{x}_\text{D} | \mathbf{z})]  \\
     &\quad+ R\,  H_B(\mathbf{x}_{\text{in}})\nonumber \\
     \nonumber\\
    & =  -  \frac{1}{2} \beta \sum_{j=1}^{m}\left(1+\log \left({\sigma}_j^2\right)-\mu_j^2-\sigma_j^2\right)\nonumber \\
    &-          
   \log {\mathcal N}_{\mathbf{x}_{\text{in}}}( \mathbf{x}_\text{D}, \text{diag}(\boldsymbol{\sigma}_\text{D})^2)
    -R  \log \left(1-p\left(\mathbf{x}_{\text{in}}\right)\right) \;.
     \label{eqn:TS}
\end{align}
The hyperparameter $R$ controls the relative importance of the binary cross-entropy term, and its optimization is discussed in ~\ref{appendix:R}.

 {In order to determine the $TS$ distribution under $\mathcal{H}_0$, a set of $10^4$ ER and $10^4$ NR events are simulated according to their expected rates after trigger-level cuts, fiducialisation and signal region cuts, as given in Fig.~\ref{fig:bkgs} of \ref{sec:bkg}. In Fig.~\ref{fig:p_dataset} we show a dataset comprised of each background component (dark/light grey histogram) as well as two injected WIMP signals (color curves) at a relatively large cross-section (for illustration purposes) in $TS$ space, re-weighted to an exposure of 200 ty. The spectral dependence of the ELBO manifests in $TS$ space, with anomalous events (in this case, WIMPs) being mapped to larger $TS$ values than the background. We therefore observe two bumps in the $TS$ distribution of the NR and ER backgrounds corresponding to the classifier's prediction. ER events that present with higher $TS$ values typically have lower energies, as would make qualitative sense due to low-energy ER being indistinguishable from NR. In \ref{appendix:encoding}, we demonstrate that the VAE non-trivially encodes the spectral energy information of all events (both NR and ER), despite the VAE having been trained only on ER events.}

\subsection{Neural networks training and validation}
\label{sec:data}

The neural networks are trained on vectorized formats: [\texttt{S1WaveformTotal}, \texttt{S2WaveformTotal}, \texttt{S2Patterns}], with a total size of 3835. The waveform and hit pattern data provide information about each event, making it possible for the neural anomaly detector to learn complex features pertaining to the class of the event (ER vs NR) as well as the different spectral dependency of each class (see \ref{appendix:encoding} and \ref{appenix:VAE} for further details).

We generate training data sets consisting of an even sample of $2\times 10^4$  ER and NR events with true recoil energies uniformly distributed in $E_R\in[1,100]$ keV, with 30\% being kept aside for validation.  The average training time per epoch is $\sim$ 1 second for the VAE ($\sim40$ seconds total training time) and $\sim0.8$ seconds for the classifier ($\sim8$ seconds total training time) on an NVIDIA A100-PCIE-40GB GPU. Testing times event-by event are of the order of ms.

\subsection{Null hypothesis test}
In order to test for the presence of an anomalous bump (due to anomalous, non-background-like events) in the $TS$ distribution, we define an unbinned 1D likelihood for the background probability distribution function (pdf), $f_0$, called the `extended Poisson' \cite{ParticleDataGroup:2022pth}:
\begin{align}
    \mathcal{L}(\mathbf{TS}|\mathcal{H}_0 ) =   \frac{e^{-B}}{N!}\prod_{i=1}^NB f_0\left(TS_i \right) \, .
    \label{eqn:likelihood}
\end{align}
Here \(\mathbf{TS}\) denotes the vector of observed  $TS$ produced by the trained neural network for events labeled by $i$ during a given exposure, while \(B\) is the total expected number of background events and $N$ is the number of observed events. 
\begin{figure}[!t]
    \centering
    \includegraphics[width=.5\textwidth]{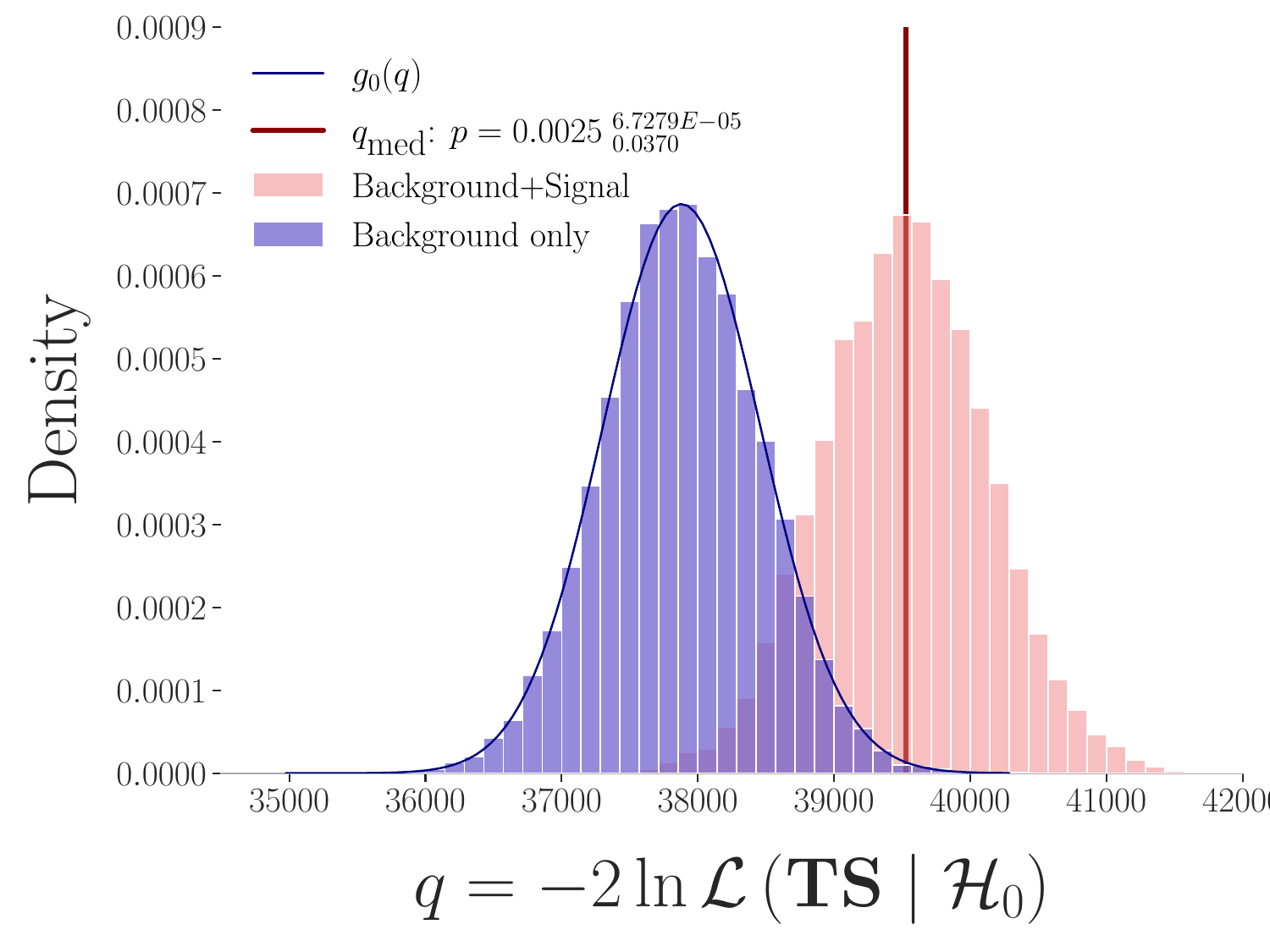}
\caption{ {Distribution of $q=-2 \ln \mathcal{L}(\mathbf{T S} \mid \mathcal{H}_0)$ from pseudodata generated under $\mathcal{H}_0$ (blue) and with an injected dark matter (WIMP) signal with $\sigma_\textrm{SI}=6.5\times10^{-48}$ cm$^2$ and $m_\chi = 50$ GeV (pink), which yields a median sensitivity of $\sim3\sigma$ at 200ty exposure. We also display as a blue line the kernel density estimate (KDE) used to evaluate the integral in Eq.~\eqref{eq:pval_med}. The red vertical line denotes $q_\textrm{med}$.}}
    \label{fig:med_ex}
\end{figure}

We take as a test statistic the distribution of $q = -2\ln\mathcal{L}$, formalizing $\mathcal{H}_0$ as the asymptotic distribution of $q$ after simulating $\sim10^4$ experiments, each with an exposure of 200~ty, using pseudo-datasets comprised solely of background events, where the number of events per experiment is sampled from a Poisson with expectation value $B$, leading to a number of events per experiment $\sim\mathcal{O}(6.5\times10^3)$. This distribution of $q$ is shown in blue in Fig.~\ref{fig:med_ex}. Any upward fluctuation of the negative log-likelihood denotes a departure from the background-only hypothesis by construction. The distribution of $q$ from another $10^4$ simulated experiments including an injected WIMP signal at a fixed benchmark of  $\sigma=6.5\times10^{-48}$cm$^2$, $m_\chi = 50$ GeV is shown in pink, while the median significance $q_\textrm{med}$ (i.e., the median  $p-$value for which one can reject $\mathcal{H}_0$ in the presence of a signal, calculated over a collection of pseudo-datasets~\cite{Cowan:2010js}) is denoted by the vertical red line. The median sensitivity is the $p-$value to reject $\mathcal{H}_0$ corresponding to $q_\text{med}$:
\begin{align}
p_\text{med} = \int_{q_\text{med}}^\infty \,dq\; g_0\left(q\right) \;, 
\label{eq:pval_med}
\end{align}
where $g_0(q)$ is the distribution of $q$ under the null hypothesis. 



\section{Results}
\label{sec:analysis}
\begin{figure*}[t]
    \centering
    \hspace{-2cm}\includegraphics[width=0.49\textwidth]{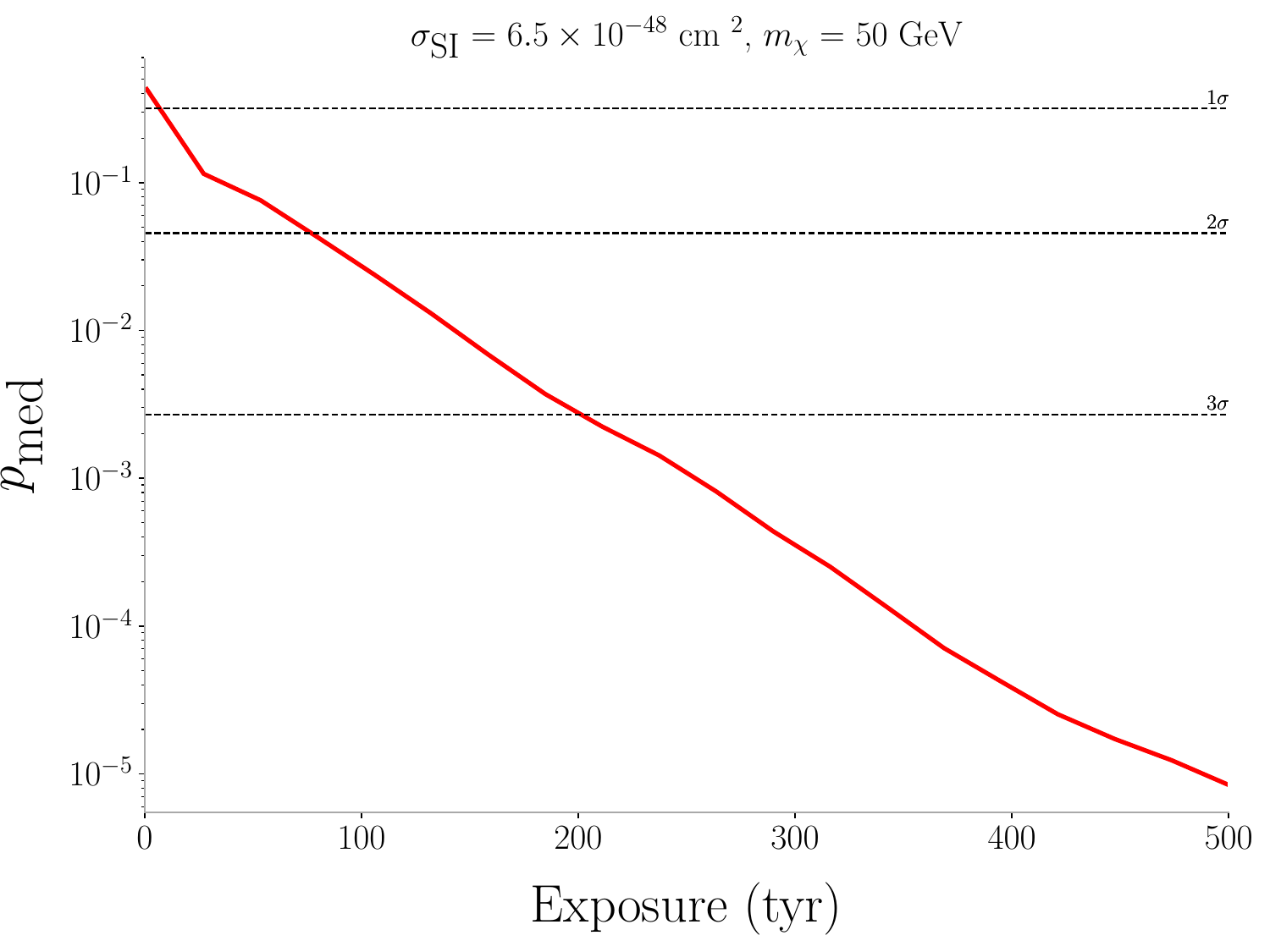}\includegraphics[width=0.64\textwidth]{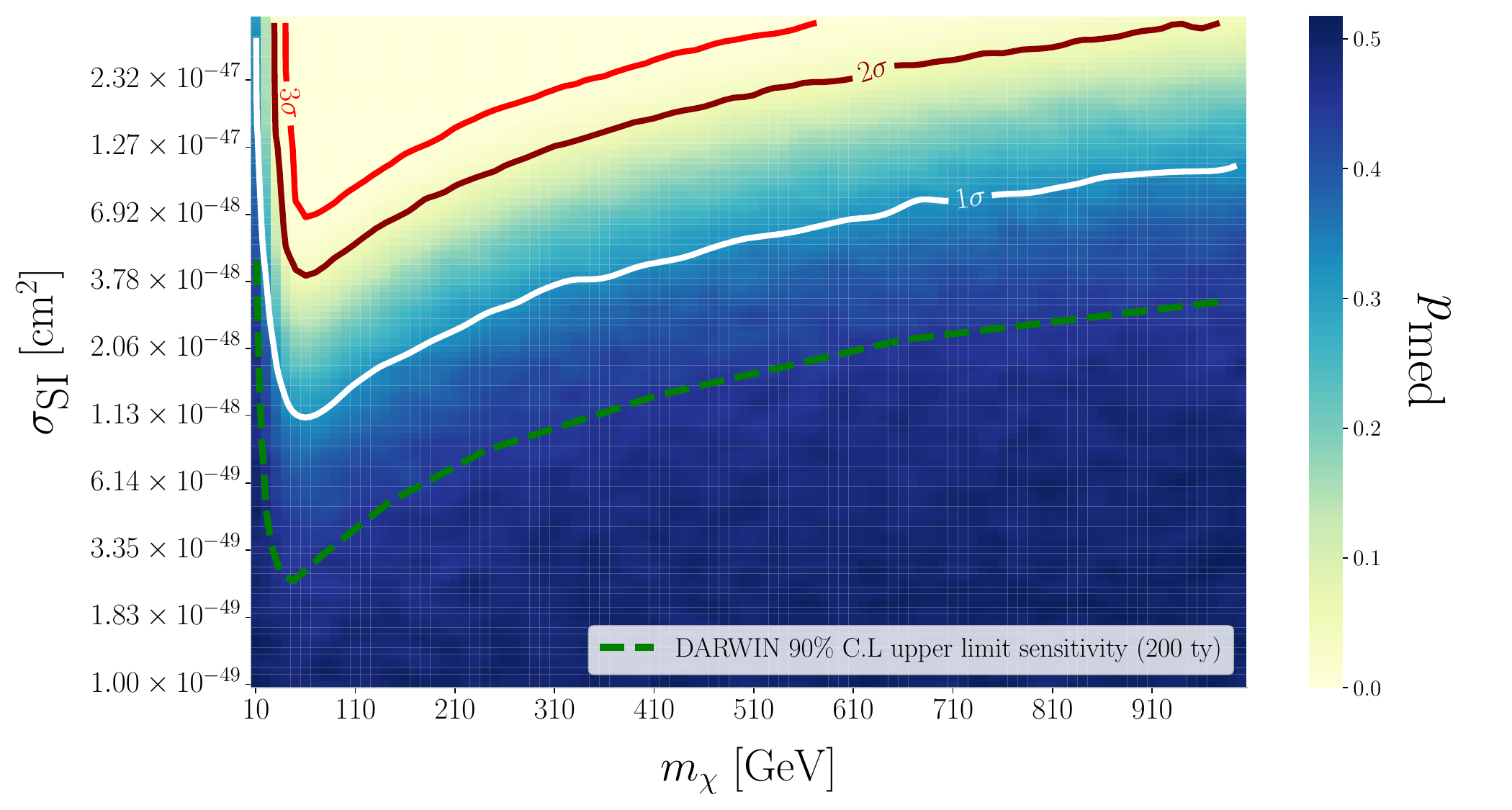}
    \caption{\textbf{Left:}  {Median sensitivity to reject the background-only hypothesis as a function of detector exposure at the benchmark $\sigma_{\mathrm{SI}}=6.5 \times 10^{-48} \mathrm{~cm}^2, m_\chi=50\, \mathrm{GeV}$. Thresholds of 1,2 and $3\sigma$ decision boundaries are shown as black horizontal dashed lines.}  \textbf{Right:} Median sensitivity in the $m_\chi$, $\sigma_\textrm{Si}$ plane from the anomaly detection pipeline (exposure of 200 ty), with contours at $1, 2$ and $3\sigma$ (solid lines). For qualitative comparison, the WIMP-model dependent DARWIN  90\% C.L. median upper limit sensitivity is shown as the green dashed line. }
    \label{fig:prob_vs_exp}
\end{figure*}

In this section, we present the results from our approach on simulated data. For this analysis, the ER and NR background distributions have been re-weighted to their expected values using the background benchmarks from ~\ref{sec:bkg}.
The median sensitivity to reject $\mathcal{H}_0$ as a function of exposure is shown as the red line in Fig.~\ref{fig:prob_vs_exp} (left panel) for the WIMP benchmark adopted in Fig.~\ref{fig:med_ex} ($\sigma_\text{SI} = 6.5\times 10^{-48}$ cm$^2$, $m_\chi = 50$ GeV).

The right panel of Fig.~\ref{fig:prob_vs_exp} shows the median sensitivity in the canonical 2D WIMP parameter space for a fixed exposure of 200~ty. We plot the median sensitivity as a color gradient, indicating contours corresponding to  $1, 2$ and $3\sigma$ median sensitivity.  For qualitative comparison only, we display the 2016 median DARWIN 90\% C.L. upper limit sensitivity as a green dashed curve \cite{Aalbers:2016jon}. It is important to note that this 90\% C.L upper limit sensitivity is not directly comparable to the background rejection test in our pipeline, as these are two fundamentally different statistical tests: the 90\% C.L upper limit sensitivity is model-dependent (as the WIMP signal is specific for a given model), whilst the anomaly detection method is agnostic to the WIMP physics, as the neural networks were only trained on samples indicative of a background-only dataset, with no information about WIMP-like events. Hence, whilst the background rejection $p$-value we present is a somewhat `stronger' statistical claim (in that it is model-independent), we find (as expected) that an upper limit in the presence of an explicit alternative WIMP model is significantly more constraining.

\section{Conclusions}
\label{sec:conclusion}

This study presents the foundation for a deep learning analysis pipeline to perform anomaly detection in next next-generation dark matter direction detection experiment -- in this case, the DARWIN design. The proposed methodology provides a prototype for future developments in statistical inference in rare physics searches with xenon-based TPCs, and promises to extract maximal information from the high-dimensional event data produced by TPC experiments. This is particularly critical given the current challenges faced by modern TPC experiments, where a substantial portion of analysis time is devoted to tuning optimal cuts and corrections for high-level, compressed summary observables.

  {The method in this paper presents an anomaly-aware machine learning technique that leverages deep learning to conduct a background  rejection task. We use a neural network architecture consisting of an unsupervised VAE and a fully connected classifier that extracts relevant event-by-event features (including energy information) from PMT hit pattern data and total S1 and S2 waveforms.  We find that the neural anomaly detector achieves sensitivity to reject $\mathcal{H}_0$ at the order of $3\sigma$ after  $\sim 200$ ty for a WIMP benchmark of $\sigma_{\mathrm{SI}}=6.5 \times 10^{-48} \mathrm{~cm}^2, m_\chi=50\, \mathrm{GeV}$. }

A model-independent anomaly detection can serve as a `first pass'  analysis, assessing if there is any data that is not consistent with the background-only expectation, before moving on to a more sensitive model-dependent search (e.g., via likelihood ratio). Whilst we have validated our pipeline in the context of a canonically interacting WIMP, the machinery remains identical for any new physics search. This makes the development and deployment of these types of analyses an important addition to the standard statistical pipeline. 

 {As is always the case for simulation-based analyses, the neural networks could be subject to missing or misinterpreting key underlying features or stochastically of real data should the simulations be incomplete or otherwise imperfect \cite{hermans2021averting,edwards1985exact}. To mitigate this risk, one could expand the pipeline to include fine-tuning the models on calibration data in the training of the neural network, thereby complementing simulated events with actual observations. A large computational effort is currently being directed toward folding in calibration information into the derivation of the high-level cS1/cS2 statistics, something that would  {be complemented by} our approach: a neural network-based analysis pipeline can alleviate the computational burden as it bypasses the need for these corrections. However, care must be taken with uncertainties due to specification of the recoil energy of events, especially at lower energy thresholds \cite{akerib2016lowenergy,lenardo2019ionization}. This type of issue could be circumvented with unsupervised anomaly detector networks that have integrated domain adaptation between simulated source data and target calibration \cite{gong2018causal}. Investigation of these types of models will be the subject of future work.  }  

Given the simulation-rich environment at DARWIN and in the future, XLZD, we plan to leverage this approach, including multi-scatter classification, energy and position reconstruction, circumventing the need for traditional detector fiducialisation or signal region definition. Other architecture developments will be aimed at handling high-dimensional temporal PMT data, accidental coincidence, and surface events background discrimination, as well as inter-ER background classification.

\section{Acknowledgements}
AS was partially supported by the grant ``DS4ASTRO: Data Science methods for Multi-Messenger Astrophysics \& Multi-Survey Cosmology”, in the framework of the PRO3 `Programma Congiunto' (DM n. 289/2021) of the Italian Ministry for University and Research. RT and AS acknowledge funding from Next Generation EU, in the context of the National Recovery and Resilience Plan, Investment PE1 – Project FAIR ``Future Artificial Intelligence Research''. This resource was co-financed by the Next Generation EU [DM 1555 del 11.10.22]. RT is partially supported by the Fondazione ICSC, Spoke 3 ``Astrophysics and Cosmos Observations'', Piano Nazionale di Ripresa e Resilienza Project ID CN00000013 ``Italian Research Center on High-Performance Computing, Big Data and Quantum Computing'' funded by MUR Missione 4 Componente 2 Investimento 1.4: Potenziamento strutture di ricerca e creazione di ``campioni nazionali di R\&S (M4C2-19 )'' - Next Generation EU (NGEU).
This work was also supported by the Swiss National Science Foundation 
under grants No 200020-162501 and No 200020-175863, by the European Union’s Horizon 2020 
research and innovation programme under the Marie Sklodowska-Curie grant 
agreements No 674896, No 690575 and No 691164, by the European Research Council (ERC) 
grant agreements No 742789 (Xenoscope) and No 724320 (ULTIMATE), by the Max-Planck-Gesellschaft, 
by the Deutsche Forschungsgemeinschaft (DFG) under GRK-2149, by the US National Science Foundation (NSF) 
grants No 1719271 and No 1940209, by the Dutch Science Council (NWO), by the PortugueseFCT, 
by the Ministry of Education, Science and Technological Development of the Republic of Serbia and 
by grant ST/N000838/1 from Science and Technology Facilities Council (UK). We further acknowledge funding from the German Federal Ministry for Research, Technology and Space (BMFTR) and from the Helmholtz Association.
\input{appendix.tex}

\bibliographystyle{spphys}             
\bibliography{DLDD_unsup_PAPER}   
\end{document}

%% file: appendix.tex
\appendix
\section{Background modeling}

\begin{figure*}[t]
    \centering
\includegraphics[width=0.5\textwidth]{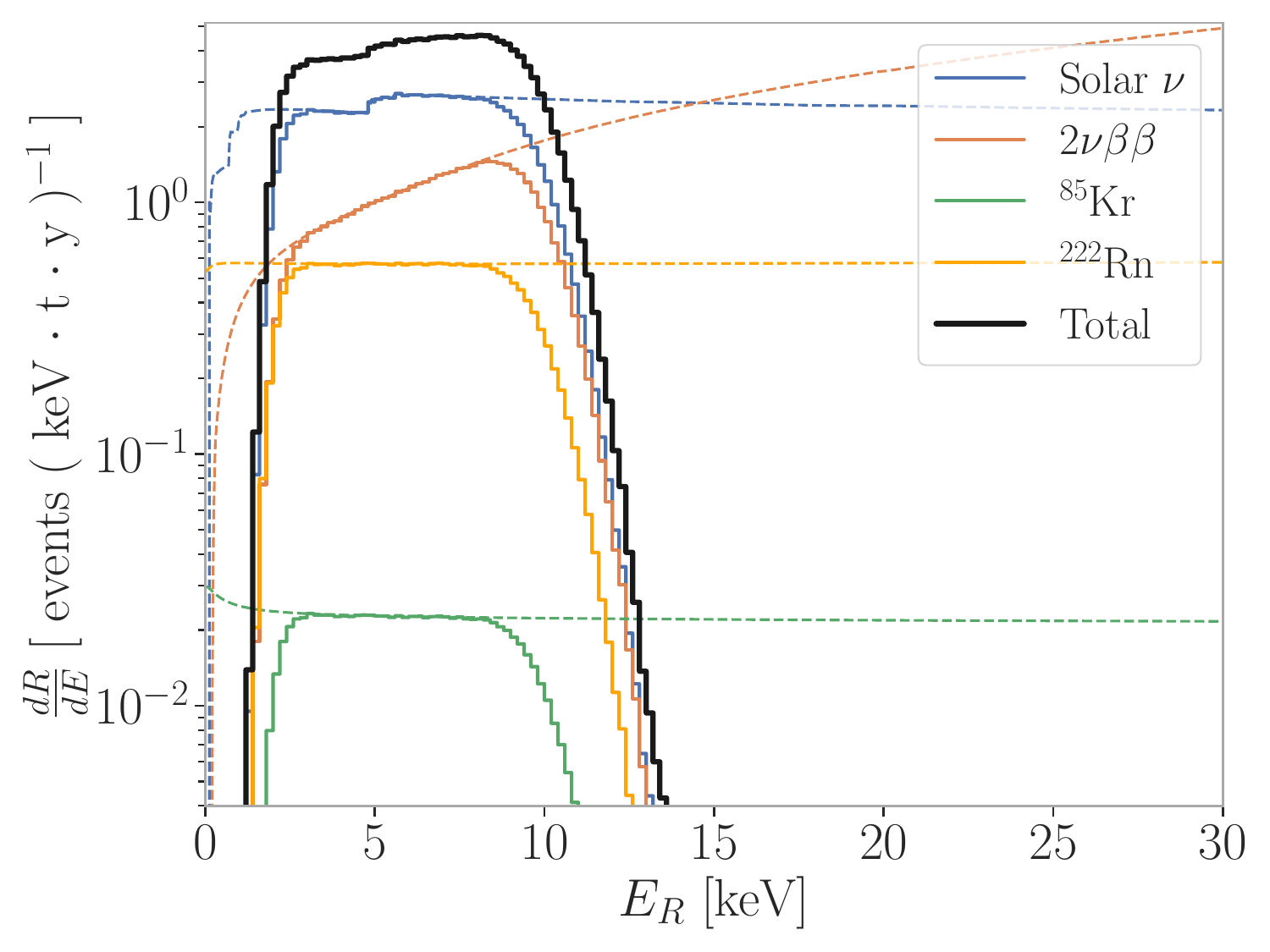}~ \includegraphics[width=0.5\textwidth]{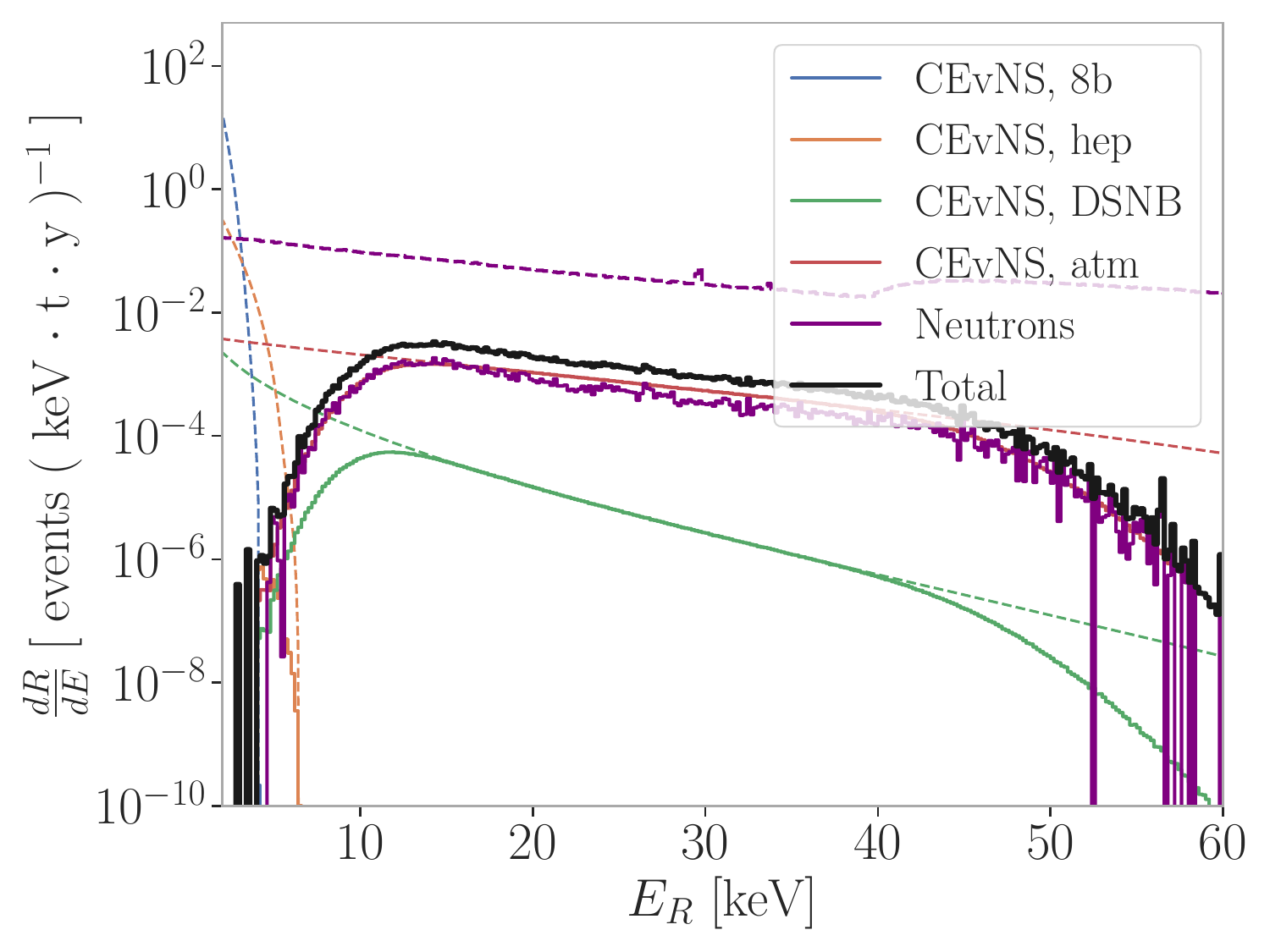}   \caption{ Benchmark DARWIN background differential recoil rate spectra considered in this analysis, before (dashed lines) and after (solid lines) detector-level SR, fiducialization and threshold cuts. The total background contributions are shown by black solid lines.  \textbf{Left:} ER backgrounds originating from low-energy solar neutrinos, two-neutrino double-beta decays of $^{136}$Xe and intrinsic backgrounds from $^{85}$Kr and $^{222}$Rn. \textbf{Right:} NR background contributions, produced by coherent neutrino-nucleus scattering sources: solar neutrinos originating from $^8$B and from the helium-proton reaction, atmospheric neutrinos, the diffuse supernova neutrino background, and radiogenic neutrons from the detector. The mean integrated rates are given in Table~\ref{tab:n_uncert} for each background component.}
    \label{fig:bkgs}
\end{figure*}
\label{sec:bkg}
{In this section, we detail the background modeling of this study}. The different sources of ER and NR backgrounds relevant to DARWIN are described in Refs.~\cite{DARWIN:2023uje,Schumann_2015,Aalbers:2016jon}.

\begin{table}[t]
    \centering
    \begin{tabular}{|l|c|}
        \hline
        \multicolumn{2}{|c|}{\textbf{Background event rates}} \\ \hline
        & \textbf{$\overline{\mathrm{Rate}}\;\big[(\mathrm{ty})^{-1}\big]$} \\ \hline
        ER intrinsic: $^{136}$Xe $(2\nu\beta\beta)$ & 9.4 \\
        ER intrinsic: $^{222}$Rn                    & 4.5 \\
        ER intrinsic: $^{85}$Kr                     & 0.18 \\
        ER solar neutrinos                          & 20.0 \\ \hline
        NR solar CEvNS                              & $5.3\times10^{-4}$ \\
        NR atmospheric CEvNS                        & $2.6\times10^{-2}$ \\
        NR radiogenic neutrons                      & $2.2\times10^{-3}$ \\ \hline
    \end{tabular}
    \caption{Summary of mean background event rates after detector-level SR, fiducialization and threshold cuts.}
    \label{tab:n_uncert}
\end{table}

The background contributions in DARWIN can be categorized into external and intrinsic backgrounds: external backgrounds include gamma-rays and neutrons originating from radioactive decays or interactions outside of the target volume. These can be significantly reduced by target fiducialization due to the high density of liquid xenon. Intrinsic backgrounds, on the other hand, are uniformly distributed in the target region and cannot be reduced by fiducialization\footnote{In this study, we neglect surface events \cite{xenoncollaboration2024xenonntwimpsearchsignal} and isolated light and charge signals from accidental coincidences \cite{Aalbers_2023} that were considered in the analyses of XENONnT and LZ. Modeling these backgrounds is under current development at DARWIN/XLZD, and so we leave their treatment to future work.}. 

The background is obtained after detector-level cuts, including the finite energy threshold of the detector, the fiducial region and signal region (SR) cuts on the combined energy scale (CES), and an estimate of the true deposited recoil energy, $E_R$. For this analysis, we adopt a standard value of 31.5 t \cite{Aalbers:2016jon}, using an estimated location in the detector for fiducialization cuts. Furthermore, given that the spectral information of all relevant backgrounds is not currently fully known, we apply a [2-10] keVee cut on the CES of each event in line with previous studies~\cite{Schumann_2015}. This leaves ERs with a ground truth $E_R$ between $\sim$[2-14] keV and NRs between $\sim$[2-60] keV, which are displayed in Fig.~\ref{fig:bkgs}. A further assumption we make is that multi-scatter events are fully vetoed. Thus, the analysis presented in this work assumes 100\% single-scatter selection efficiency\footnote{Work is being conducted to incorporate multi-scatter selection using deep learning to supplement the pipeline presented in this work.}. {Each background contribution has an expected rate as shown in Table~\ref{tab:n_uncert}.}

 \textbf{ER backgrounds:} Solar neutrinos produced through the proton-proton ($pp$) fusion process and the subsequent beryllium-7 ($^{7}$Be) reaction in the Sun are the dominant source of ER background for dark matter searches beyond the ton-scale. This is because of their relatively low energies and high abundance, along with the fact that their contribution cannot be reduced by target purification, fiducialization, nor single-scatter selection. 
 Intrinsic backgrounds, including contributions from isotopes such as \textsuperscript{85}Kr, a beta-emitter present in natural krypton, and \textsuperscript{222}Rn, are included. These intrinsic backgrounds are uniformly distributed in the target due to the chemical inertness of noble gases.  Two-neutrino double-beta decays ($2\nu\beta\beta$) of \textsuperscript{136}Xe yield a background that steeply rises with recoil energy.
 Finally, ER backgrounds originating from  $\gamma$-rays from radioactive contamination in the cryostat and detector materials are reduced to negligible amounts by target fiducialization, hence we neglect them here~\cite{Schumann_2015}. 
 The differential energy spectra of the above four ER background contributions are shown in Fig.~\ref{fig:bkgs} (left panel), both before and after detector-level event cuts.  
\newline
\textbf{NR backgrounds:} Radiogenic neutrons emitted from the detector's materials, particularly from light PTFE used as insulator and light reflector, and photosensors made from various materials constitute a primary source of NR background\footnote{Work is currently being undertaken to improve the understanding of radiogenic neutrons in DARWIN as well as the uncertainty on their contribution. The resulting insights could very easily be included in the pipeline presented in this work in a future iteration.}. Fiducialization of the detector volume serves as the primary detector-level cut on the radiogenic neutrons, which extensive \texttt{Geant4} simulations indicate as interacting primarily near the detector walls. Furthermore, neutrons can scatter multiple times within the detector volume. A  veto on such multi-scatter events, determined from the S2 area distribution, is implemented with an assumed 100\% efficiency.   

The neutron background contributes more at larger (10-50 keV) recoil energies relative to the significantly more perilous other NR backgrounds, namely, 
 coherent elastic neutrino-nucleus scattering (CEvNS)~\cite{Schumann_2015}. $\,^8$B solar neutrinos are primarily responsible for a steep rise in background events at low recoil energy, hindering the detection of low-mass WIMPs (5-8 GeV). This background is difficult to distinguish from WIMP signals and represents a limit on sensitivity \cite{OHare:2016pjy}, at least for non-directional DD experiments. 

At higher recoil energies, the main CEvNS background is from atmospheric neutrinos (atm), with smaller contributions from solar neutrinos from the helium-proton reaction (hep) and the diffuse supernova neutrino background (DSNB)~\cite{DARWIN:2023uje,Strigari_2009}. The spectra of NR backgrounds considered in this study are shown in Fig.~\ref{fig:bkgs} (right panel).

Lastly, as noted in Sec.~\ref{sec:anom_detec_intro}, by propagating nuisance-parameter variations through the simulation and training on the resulting samples, the neural network effectively learns to marginalize over these uncertainties. For the generation of the PMT and waveform data used in this study, such nuisance parameters included within the $\texttt{Tray}$ framework include PMT-quantum efficiencies, Light collection efficiency (LCE) uncertainties, systematic uncertainties associated with the S1/S2 reconstruction, as well as single photo-electron (SPE) area response.

\section{Neural network architectures}

\label{sec:arch}
In this section, we describe the components of the neural networks in detail. All neural networks are trained with \texttt{Tensorflow v2.15.0} \cite{tensorflow2015-whitepaper}.

\subsection{Variational Autoencoder (VAE)}
\label{sec:VAE}
\begin{table}[ht!]
\centering
\begin{tabular}{|l|l|}
\hline
\multicolumn{2}{|c|}{\textbf{Variational Autoencoder Architecture}}                       \\ \hline
Latent Dimension, $m$               & 128                                                       \\ \hline
$\beta$              & 10                                                        \\ \hline
{Encoder}       & Input Layer: Shape: 3835                               \\
                               & Dense Layer: 2000 units                                   \\
                               & Dense Layer: 500 units                                    \\
                               & Dense Layer: $2 \times $m                         \\ \hline
{Decoder}       & Input Layer: Shape: $m$                    \\
                               & Dense Layer: 500 units ($\times 2$)                               \\
                               & Dense Layer: 2000 units ($\times 2$)                              \\
                               & Dense Layer: 3835 units ($\times 2$)                              \\ \hline
Optimizer                      & Adamax, Learning Rate: 0.0005                             \\ \hline
Training Epochs               &  30                                                \\ \hline
\end{tabular}
\caption{Summary of the VAE architecture and optimal hyperparameters.   {All dense layers have ReLU activations except for the linear input and output.} }
\label{tab:vae_architecture}
\end{table}

The goal of an autoencoder is to learn a compressed representation (encoding) of the input data, and then reconstruct the input data from this encoding~\cite{bank2021autoencoders,Schmidhuber_2015}. Autoencoders encompass three primary components: an encoder, a latent space, and a decoder. 
The encoder reduces the input data vectors $\textbf{x}_\text{in} \in \mathbb{R}^n$ into a lower-dimensional latent space representation $\textbf{z} \in \mathbb{R}^m$ (with $m\ll n$) through a transformation $\textbf{z}=f(\textbf{x})$. The decoder then reconstructs the input from this compressed form, aiming to produce an output $\textbf{x}_D =g(\textbf{z})$ as close to the original $\textbf{x}_\text{in}$ as possible. A reconstruction loss function, quantifying the difference between $\textbf{x}_\text{in}$ and $\textbf{x}_D$, is optimized during training.

Variational Autoencoders (VAEs)  extend this concept by introducing a probabilistic approach to the encoding process. Unlike standard autoencoders, the encoder in a VAE maps input data to a probability distribution characterized by mean $\mu$ and variance $\sigma^2$, essentially transforming the encoder’s output into the parameters of a Gaussian distribution:
$$
f(\textbf{x}_\text{in}) 
\rightarrow q(\textbf{z} \mid \textbf{x}_\text{in})
= {\mathcal N}_\textbf{z}\left({\boldsymbol{\mu}, \text{diag}(\boldsymbol{\sigma}^2)}\right) \, .
$$
The decoder, now governed by $g(\mathbf{z})\rightarrow p(\mathbf{x}_D|\mathbf{z})$, is a probabilistic distribution that reconstructs data from sampled points in this probabilistic latent space. When $\mathbf{x}_D$ are real vectors, $p(\mathbf{x}_D|\mathbf{z})$ is taken to be a multidimensional normal distribution with diagonal covariant structure\footnote{This is a simplifying choice for the covariance structure. See Ref.~\cite{Dorta2018Training} for an application of a structured Gaussian as the decoder. } \cite{kingma2022autoencoding}:
\begin{align}
   p(\mathbf{x}_D|\mathbf{z}) =  {\mathcal N}_{\mathbf{x}_{\text{in}}}( \mathbf{x}_\text{D}, \text{diag}(\boldsymbol{\sigma}_\text{D}^2))\;.
\end{align}
 
The VAE is trained via stochastic gradient descent by maximizing the loss function given by the so-called `evidence lower bound' or $\text{ELBO}$~\cite{kingma2022autoencoding}: 


\begin{flalign}
   \quad\text{ELBO} &=  \mathbb{E}_{q(\mathbf{z}|\mathbf{x}_{\text{in}})}[\log p_{\mathbf{x}_{\text{in}}}(\mathbf{x}_\text{D} | \mathbf{z})] - D_\text{KL}(q(\mathbf{z} | \mathbf{x}_\text{in}) || p(\mathbf{z}))  \nonumber\\
   &=\frac{1}{L} \sum_{l=1}^{L} 
   \log {\mathcal N}_{\mathbf{x}_{\text{in}, l}}( \mathbf{x}^\text{D}_l, \text{diag}(\boldsymbol{\sigma}^\text{D}_{l})^2)\nonumber \\
   &\quad\quad\quad+  \frac{1}{2} \beta \sum_{j=1}^{m}\left(1+\log \left({\sigma}_j^2\right)-\mu_j^2-\sigma_j^2\right) \;,
    \label{eqn:recon_loss}
\end{flalign}
where $m$ is the dimensionality of the latent space (number of independent Gaussians), the expectation is under the distribution $q(\textbf{z} \mid \textbf{x}_{\text{in}})$ and the data are batched into batches of size $L$. The coefficient $\beta$ in the Kullback-Leibler (KL) term balances its regularization strength~\cite{sikka2019closer}, with a higher $\beta$ value ensuring that the encoded representations are closer to the prior, $p(\textbf{z})$, taken to be a standard multivariate Gaussian, $\mathcal{N}({\bf 0}_m,{\bf 1}_m)$. 


 The VAE architecture used in this study was selected after hyperparameter optimization on validation datasets withheld from training, and inspired by previously successful architectures in similar settings, in particular Ref.~\cite{us_sup}. It consists of an encoder that takes vectorized data inputs $\mathbf{x}_\text{in}$  (see Sec.~\ref{sec:data}) in batches of size $L=10$ and processes it through two dense (i.e., fully-connected) layers with 2000 and 500 units respectively. The latent space dimension is $m=128$. The decoder has a dual-network structure. Both networks within the decoder begin with an input of shape 128, and process it through dense layers of 500 and 2000 units, culminating in two output layers $\mathbf{x}_D$ and $\log \boldsymbol{\sigma}^2$ with shape matching $\mathbf{x}_\text{in}$. The architecture is summarized in Table~\ref{tab:vae_architecture}. We note that this architecture does not scale for use on raw time series PMT readout data, given that dense, fully connected neural networks are not optimal for the sparsity one would expect from such data (leading to optimization issues and computational inefficiency). Therefore, a more suitable architecture, such as the one presented in Ref.~\cite{arxiv_sim_inf}, would be needed to use raw data as input. 
 
For training, we use an Adamax optimizer with a learning rate of \(0.5 \times 10^{-3}\). The entire training regimen is set to run for 30 epochs, with an optimized $\beta$ value of 10 (via uniform hyperparameter scans). Validation tests of the so-obtained latent space representation are presented in ~\ref{appenix:VAE}.


\subsection{Supervised ER vs NR Classifier}
\label{appendix:classifer}

\begin{table}[hb]
\centering
\begin{tabular}{|l|l|}
\hline
\multicolumn{2}{|c|}{\textbf{Classifier Architecture}}              \\ \hline
Input Shape                   & Data Shape (3835)                               \\ \hline
{Layers}       & Dense Layer: 256 units           \\
                              & Dense Layer: 64 units           \\
                              & Dense Layer: 16 units           \\
                              & Output Layer: 1 unit         \\ \hline

Optimizer                     & Adam, Learning Rate: 0.01              \\\hline
Training Epochs               &  5                                           \\ \hline\end{tabular}
\caption{Summary of the neural network classifier's architecture and optimal hyperparameters. The activation function of all layers is ReLU, with the output being a sigmoid.}
\label{tab:nn_classifier}
\end{table}

{The second component of the anomaly detector pipeline is a simple multi-layer perceptron (MLP)  feed-forward neural network \cite{Goodfellow-et-al-2016}, whose architecture details are listed in Table~\ref{tab:nn_classifier}. The classifier's task is to perform a binary classification between ER (0 output value) and NR (output value of 1) events. The MLP is trained by minimizing the standard binary cross-entropy loss: }
\begin{align}
\label{eq:cross_ent}
H_B=-\frac{1}{L} \sum_{i=1}^L \log \left(1-p\left(\mathbf{x}_{\text{in}}\right)\right)
\end{align}           
where \( L = 10\)  is the number of samples in the batch, and $p\left(\mathbf{x}_{\text{in}}\right)$ is the predicted class probability for each sample extracted from the sigmoid output of the MLP. The architecture details of this classifier are listed in Table~\ref{tab:nn_classifier}. 

\begin{figure}[t!]
    \centering
    \includegraphics[width=0.5\textwidth]{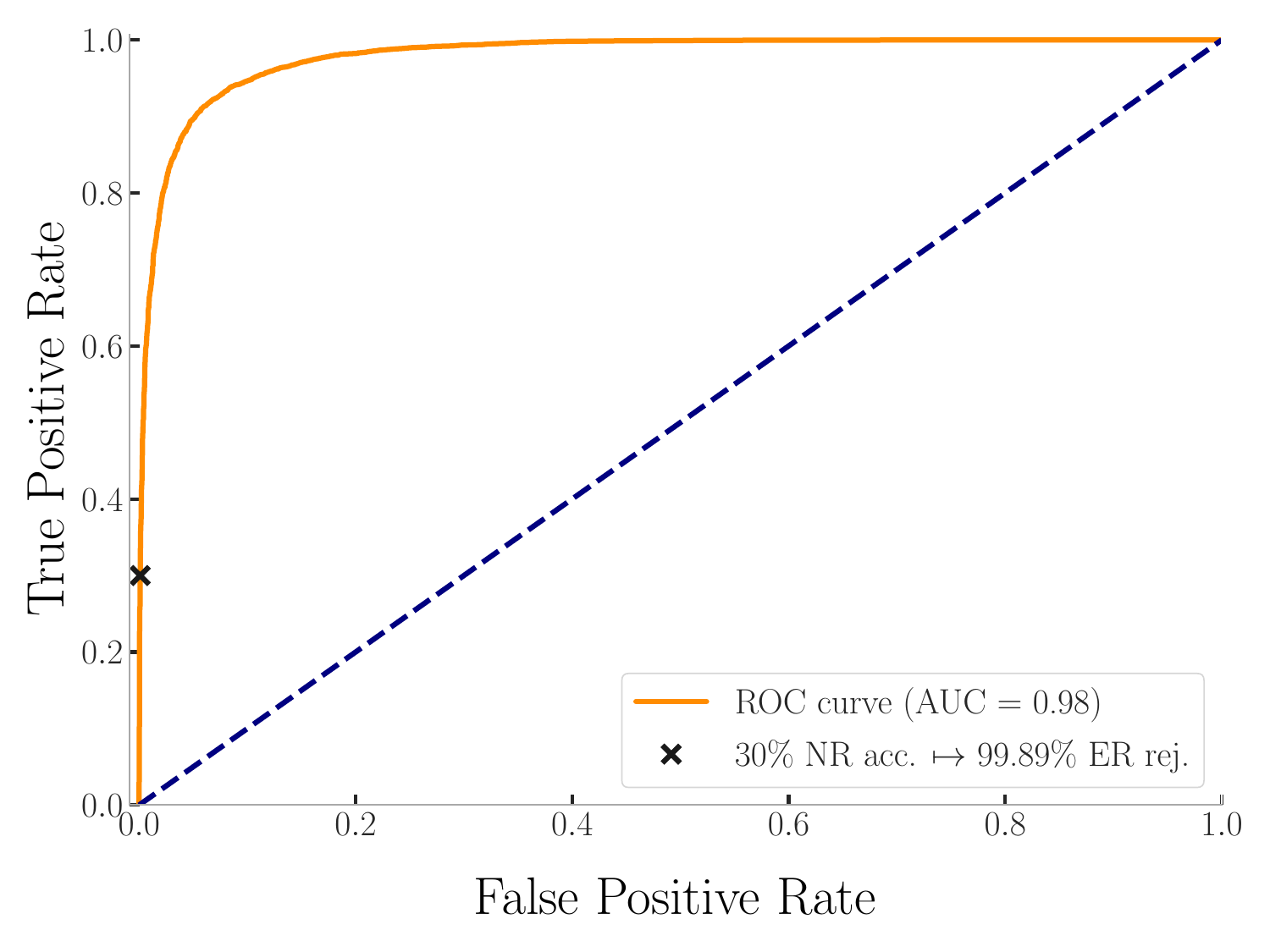}
    \caption{Receiver operating characteristic (ROC) curve of the supervised classifier trained to discriminate ER vs NR events, evaluated on a testing set consisting of an evenly mixed sample of $10^4$ NR and $10^4$ ER events. The area under the curve (AUC) is 0.98. The dashed blue lines indicate a random classifier. The black cross denotes the false positive rate (FPR) at a true positive rate (TPR) of 0.3, corresponding to the ER rejection capability of the classifier when the NR acceptance is 30\%.   }
    \label{fig:ROC}
\end{figure}

{In Fig.~\ref{fig:ROC}  we show the receiver operating characteristic curve (ROC) for a test set of $10^4$ ER and NR events. We observe an area under the curve (AUC) of 0.98. We compare our classifier's performance with the 99.98\% ER rejection obtained in previous DARWIN sensitivity studies~\cite{Aalbers:2016jon,Aalbers_2023}. Such a large ER rejection probability is intended to mitigate leakage of ER's into the WIMP NR signal region, but it comes at the expense of a lower NR acceptance, which is estimated at 30\% in Ref.~\cite{Aalbers_2023}. The false positive rate (FPR) for our classifier (which corresponds to ER leakage) at a true positive rate (TPR) of 0.3 (correct NR classification), is denoted by the black cross in Fig.~\ref{fig:ROC}, and is 0.11\%, corresponding to 99.89\% ER rejection. We note that the standard approach uses the assumption of Gaussianity for cS1 and cS2 to mitigate ER leakage}.  Our classifier, however, makes no such assumption as it operates on an event-by-event basis, and hence any misclassified events will simply modify the distribution of the anomaly score, see Sec.~\ref{sec:analysis}. We also tried modifications of the loss function in Eqn.~\eqref{eq:cross_ent} to optimize the false positive rate (i.e., minimize the number of mis-classified ER). Whilst this indeed was successful, the number of mis-classified NRs also increased, leading to a net zero effect in the overall anomaly score presented in Eqn.~\eqref{eqn:TS}. Finally, we note that the choice of architecture used to perform the event-by-event ER/NR classification is largely irrelevant, as was seen in Ref.~\cite{Lopez-Fogliani:2024gzj}, where comparable performance was observed from MLPs, transformers, and boosted decision trees. 

\section{Optimization of the supervised contribution}
\label{appendix:R}
\begin{figure}[t]
    \centering
        \includegraphics[angle=-90,width=0.5\textwidth]{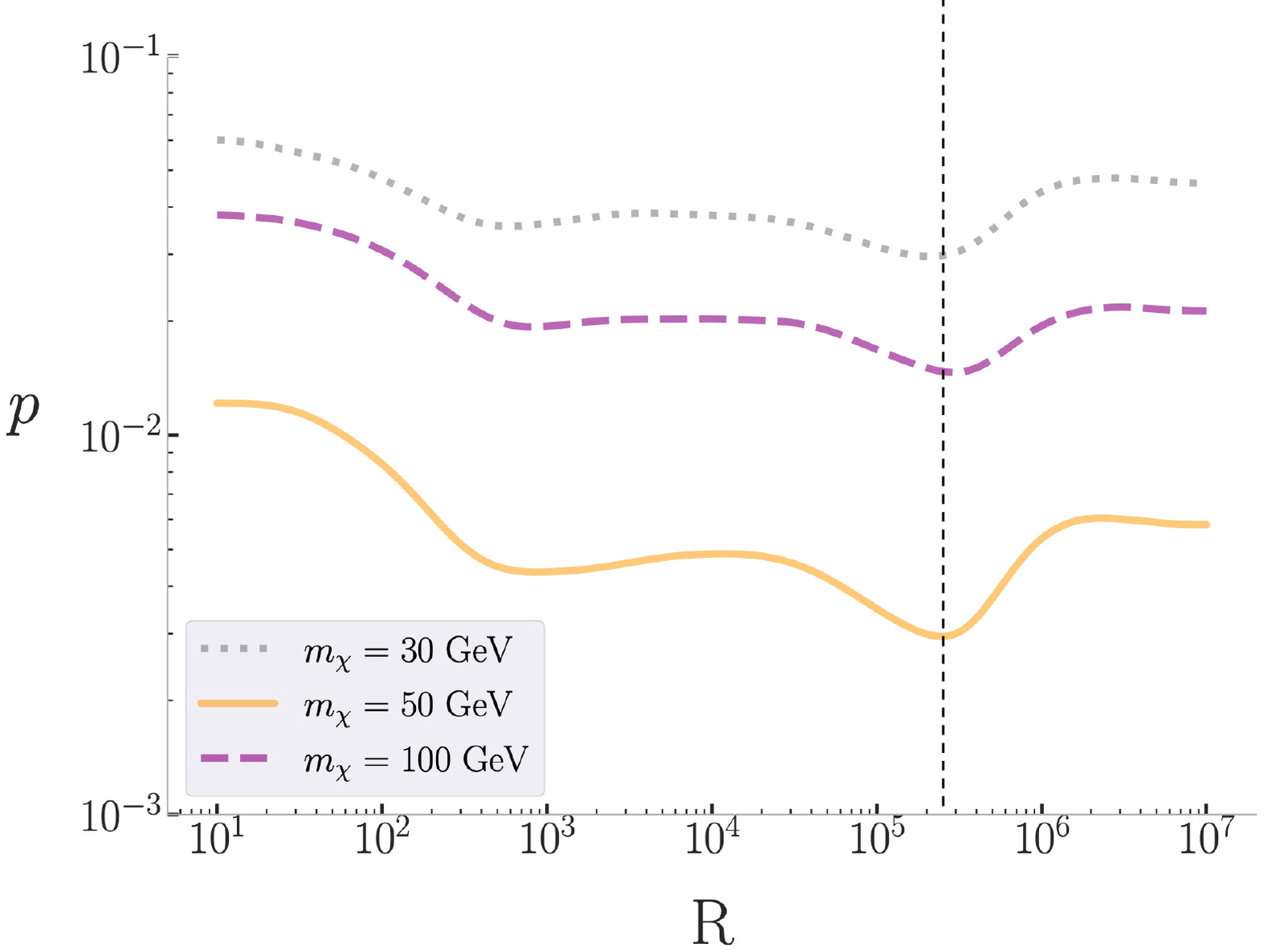}
    \caption{{optimization of the hyperparameter $R$ that controls the contribution of the supervised classifier in the determination of the anomaly score $TS$.  The $p-$value to reject $\mathcal{H}_0$ is given as a function of $R$ for three benchmark WIMP masses at fixed exposure of 200~ty and cross-section $\sigma_\textrm{SI} = 6.5\times10^{-48}$ cm$^2$. As the scattering cross-section merely rescales the median sensitivity, the choice of $R$ and cut value are insensitive to it. The optimal value for $m_\chi=50$ GeV is $R=2.5\times10^5$, shown by the vertical dashed line. The variation in the location of optimal $R$ value is small for other mass benchmarks. }   }
    \label{fig:marginal_power}
\end{figure}

The optimization of the hyperparameter $R$ in Eqn.~\eqref{eqn:TS} is a choice to be made at the time of analysis, in order to maximize the observation of any anomalous $TS$ component, if it exists. To demonstrate this, we perform a scan over a range of logarithmically spaced $R$ values $R\in[1,10^7]$ at fixed signal injection benchmarks corresponding to WIMP masses of 30, 50 and 100 GeV at an exposure of 200ty. These values of the WIMP mass were chosen in order to vary the spectral dependence of the induced WIMP signal. We show the median sensitivity, defined in  Eqn.~\eqref{eq:pval_med}, in Fig.~\ref{fig:marginal_power}, as a function of $R$, for the three benchmarks. A smaller $p-$value means better anomaly awareness and higher power to reject $\mathcal{H}_0$ in the presence of a signal,  and thus $R$ should be chosen to minimize this value.  {We conduct this test at a cross-section that yields a background rejection $p$-value of at least $\sim 3\sigma$ for $m_\chi = 50$ GeV, so as to have ample statistics to perform the test for all three mass benchmarks.} 

\begin{figure*}[!ht]
    \centering
    \includegraphics[width=0.49\textwidth]{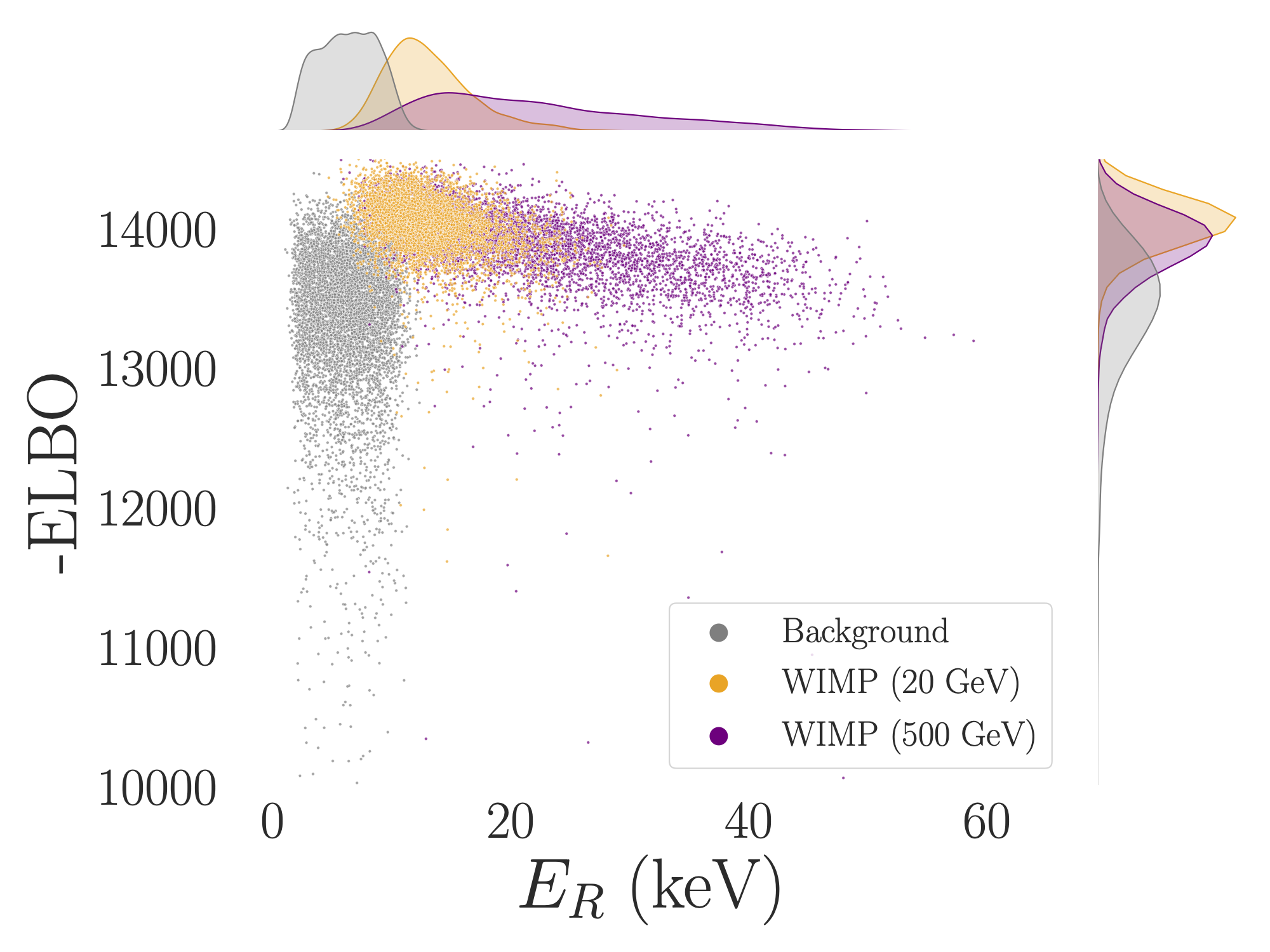}\;
    \includegraphics[width=0.49\textwidth]{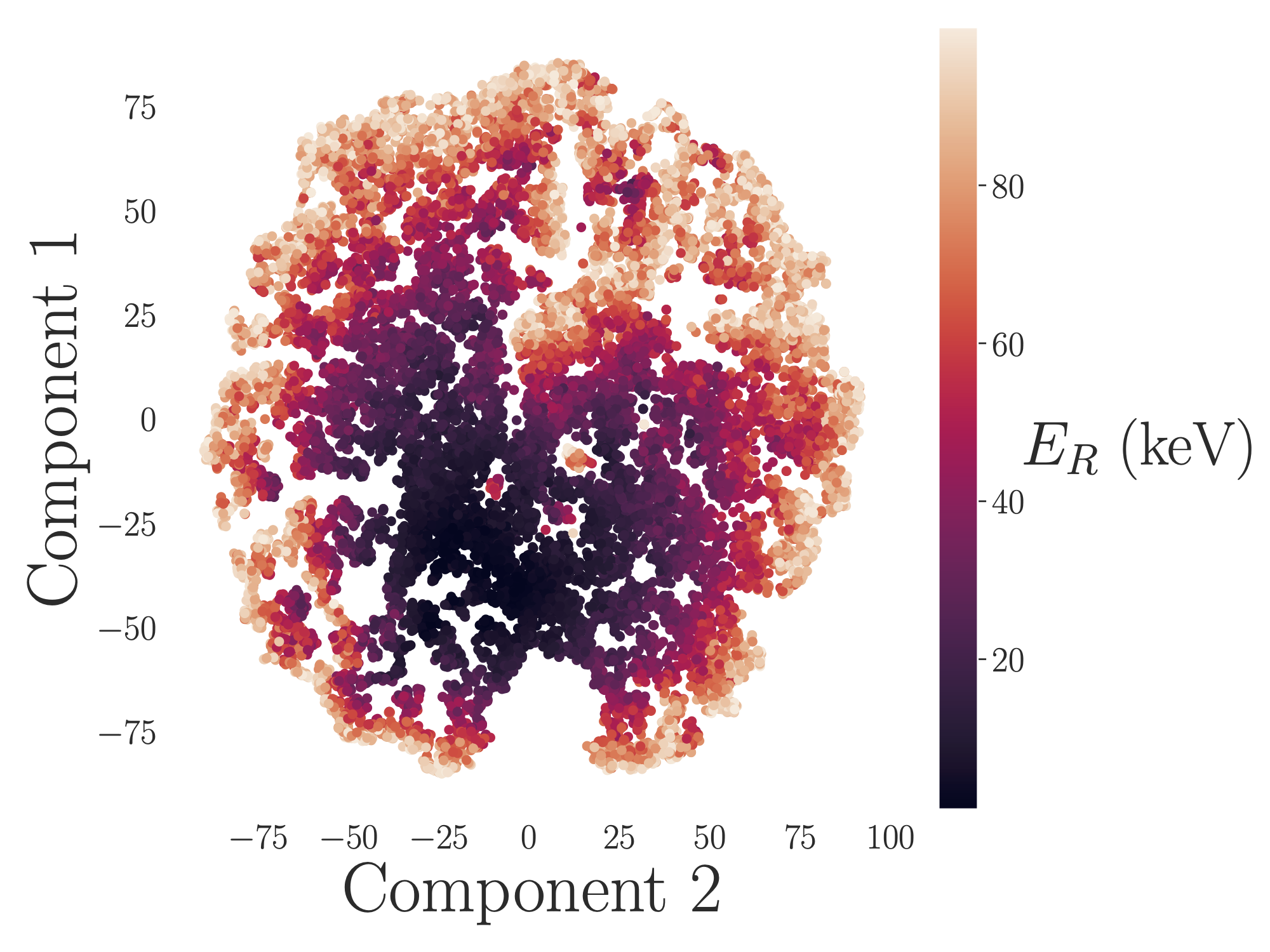}
    \caption{\textbf{Left:} ELBO values as a function of ground truth recoil energy $E_R$ for a validation set of events: total (ER+NR) background (gray) and events from two WIMP benchmarks. The 1D marginals of the ELBO and $E_R$ are also shown. {Distributions are normalized densities for illustration purposes.} The separation in the 2D space shows that spectral information has been encoded within the ELBO. \textbf{Right:} 2D tSNE of the trained VAE's 128-dimensional latent space for a validation sample of ER events with true recoil energies in the range $E_R\in[1-100]$. The color scale represents ground-truth recoil energy $E_R$ of the events. The non-trivial latent structure in $E_R$ confirms that the model has learned spectral information. 
    }
    \label{fig:tsne}
\end{figure*}

We observe that the spectral dependence of the anomaly function $TS$ entering through the ELBO as observed in Fig.~\ref{fig:tsne} does not affect the dependence of the optimal $R$ value, which lies at $\sim2.5\times10^{5}$. The general variability of the $p-$value is much more pronounced for $m_\chi=50$ GeV due to DARWINs elevated sensitivity to this mass. We observe that for $R$ values above $\sim10^6$, the $p$-value exhibits a plateau, which we have checked persists for values $R>10^7$. This indicates that above this critical value of $R$, the influence of the VAE is vanishingly small. 

The above results highlight the importance of taking a semi-supervised approach:  the fact that the power to reject $\mathcal{H}_0$ is maximized for $R\neq0, \infty$ shows explicitly the need for a combined supervised and unsupervised approach in order to maximize sensitivity to anomalous physics. In principle, $R$ could be recast as a learnable parameter during training, although we chose to leave this to future study. For this work, we adopt an optimized $R$ value of $R=2.5\times10^5$.

Previous studies observed that classifiers can perform well as anomaly detectors (see, for example, Ref. \cite{builtjes2024attention}). An admixture of many supervised and/or unsupervised components could offer additional advantages, for example, by further exploiting the topological structure of events observed in the latent space.  Indeed, Fig.~\ref{fig:marginal_power} indicates non-triviality via the two observed local minima in the $R$ dependence of the median sensitivity.  We see that the latent data feature that is learned by the VAE was the event recoil energy, whilst the classifier learns the type of event. Both of these features are crucial to a new physics discovery, regardless of origin. {It may then follow that other auxiliary models trained on the same and/or combinations/sets of prompt detector outputs may yield even better anomaly awareness. We leave this as an interesting question for future work in this domain.}
\section{Spectral information encoding}
\label{appendix:encoding}
{The distribution of the ELBO from the VAE as a function of ground truth (simulated) event recoil energy $E_R$ is shown for a validation sample with true recoil energies in the range [1-100] keV in Fig.~\ref{fig:tsne} (left panel). We plot the normalized spectral distributions in the space spanned by $E_R$ and ELBO for the total (ER+NR)  background (gray) and for events generated by two WIMP masses, $m_\chi=20,500$ GeV (orange/magenta). We observe an interesting separation in the distributions, pointing to the fact that the VAE is capable, after training exclusively on background events, to distinguish the spectral distribution from WIMPs.} 

To visualize the latent representation of the data, we further show a 2-dimensional t-distributed stochastic neighbor embedding (tSNE) projection~\cite{JMLR:v9:vandermaaten08a} of the 128-dimensional latent space of the VAE in Fig.~\ref{fig:tsne} (right panel). The non-trivial structure of the latent space, even in a two-dimensional projection, demonstrates that spectral information has indeed been incorporated into the model. 
\section{Validation of VAE representation}
\begin{figure}[t!]
    \centering
    \includegraphics[width=0.5\textwidth]{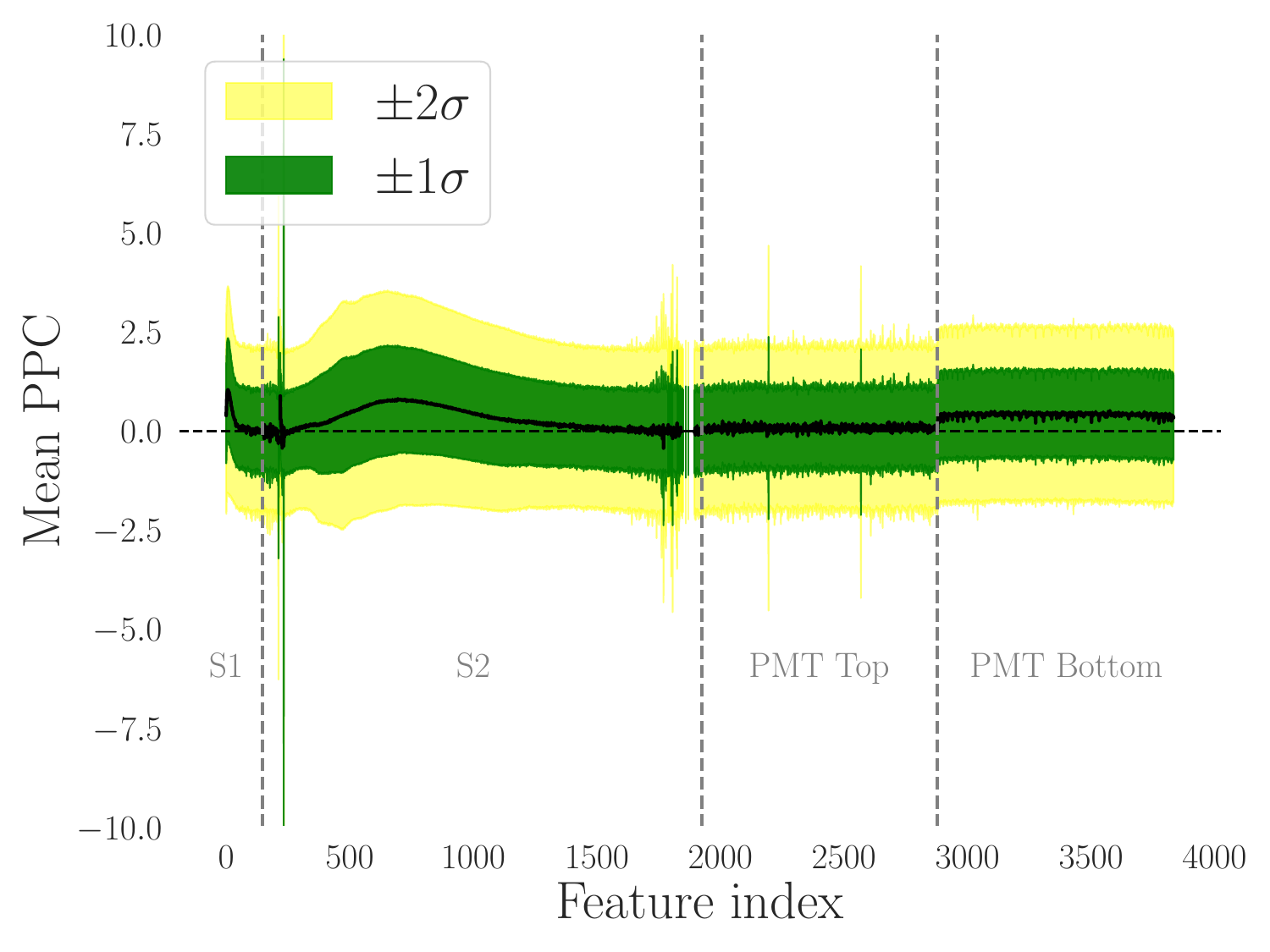}
    \caption{Posterior predictive checks performed on $10^4$ samples from the latent space of the trained VAE. A perfect VAE would produce a PPC of zero for all feature indices. The black curve is the mean PPC from  Eqn.~\eqref{eqn:PPC_mean}, with $\pm1\sigma$ and $2\sigma$ estimates shown as green and yellow bands, respectively. Each feature index corresponds to an element of the input data vector $\textbf{x}_\text{in}$. The vertical gray dashed lines demarcate the subdivision into the S1/S2 wave-forms and S2 PMT Top and PMT Bottom hit patterns.  }
    \label{fig:PPC}
\end{figure}

\label{appenix:VAE}
To validate the true low-dimensional latent features of the ER training data that were learned by the VAE, we carry out a standard benchmarking test known as a `posterior predictive check' (PPC) \cite{pritchard2015posterior,gelman2013bayesian}. {An ideal model will generatively produce samples that align with the target distribution, and therefore produce a $\text{PPC}\sim0$.} Given the one-dimensional nature of our data (after vectorisation), we adopt the following simple strategy: we generate $N$ samples ${\bf \tilde{z}} \sim \mathcal{N}({\boldsymbol{\mu}},\boldsymbol{\sigma}^2)$ from the latent space of the trained VAE and parsing them through the decoder, to obtain the predicted output, ${\bf \tilde{x}}$. A separate test set ${\bf x}_\textrm{test}$ that is withheld from training is then used to calculate the relative reconstruction error:
\begin{align} \label{eqn:PPC_mean}
    \text {Mean}\left(\text{PPC}\right)^{(j)}=\frac{1}{N}\sum_{i=1}^N\frac{\left( {{\tilde{x}}^{(j)}_{i}-{x}^{(j)}_{i_\text{test} }}\right)}{{{\sigma}^{(j)}_{\text {test } }}}\;,
\end{align}
where $\sigma_{\text{test}}^{(j)} = \sqrt{\frac{1}{N} \sum_{i=1}^N \left( x^{(j)}_{i_{\text{test}}} - \bar{x}^{(j)}_{\text{test}} \right)^2}
$ is the standard deviation of the distribution of test samples ${x}^{(j)}_{_\text{test} }$ for feature  (vector column) $j=1, \dots, 3835$ and serves as a normalization factor. We use $N=10^4$. 

We show the result of the PPC in Fig.~\ref{fig:PPC} for all 3835 data features. For clarity, we demarcate with vertical lines the features corresponding to the S1 and S2 waveforms, as well as the top and bottom S2 PMT hit-patterns. 
We plot the mean PPC as a black curve with the $\pm1,2\sigma$ uncertainties in green and yellow, respectively. While a perfect network would produce a PPC of zero for all features, we observe that our network's output lies within $1\sigma$ of 0 for all features, indicating that it has learnt the underlying properties of the training data. 
 The largest deviation from zero occurs for features at small pulse times for S1 and S2 waveforms (i.e., close to the start of the S1/S2 feature indices, indicated by the vertical lines). This is expected since most of the events used during training have a small or zero S1/S2 value at larger times (cf. Fig.~\ref{fig:data_example}). Therefore, the network has fewer issues learning this degeneracy at large times and can reconstruct the corresponding features toward the ends of the S1/S2 feature index. This, however, can lead to larger variance in the PPC distribution of some features due to the model's lack of reconstruction power in regions of degenerate zeros in the feature space, as are observed as spikes in the $1/2\sigma$ bands.  We observe near-perfect reconstruction for the top S2 PMT array, but observe a slight, positive offset for the bottom PMT. We attribute this behavior to the bottom PMT displaying what is mostly uniform noise for the majority of ER events, as seen in Fig.~\ref{fig:data_example}. Hence, the values for which the VAE can optimize the ELBO are somewhat arbitrary and present as a systematic offset. The top PMT array, however, displays concentrated deposits that are well associated with the event properties and can therefore be learned more easily.

%% file: DLDD_unsup_PAPER.bib
@article{pritchard2015posterior,
author = {David Mimno  and David M. Blei  and Barbara E. Engelhardt },
title = {Posterior predictive checks to quantify lack-of-fit in admixture models of latent population structure},
journal = {Proceedings of the National Academy of Sciences},
volume = {112},
number = {26},
pages = {E3441-E3450},
year = {2015},
doi = {10.1073/pnas.1412301112},
URL = {https://www.pnas.org/doi/abs/10.1073/pnas.1412301112},
eprint = {https://www.pnas.org/doi/pdf/10.1073/pnas.1412301112},
}

@ARTICLE{2025arXiv250520535Z,
       author = {{Zivanovic}, Uros and {Di Gioia}, Serafina and {Scaffidi}, Andre and {de los Rios}, Mart{\'\i}n and {Contardo}, Gabriella and {Trotta}, Roberto},
        title = "{Rotary Masked Autoencoders are Versatile Learners}",
      journal = {arXiv e-prints},
         year = 2025,
        month = may,
          eid = {arXiv:2505.20535},
        pages = {arXiv:2505.20535},
          doi = {10.48550/arXiv.2505.20535},
archivePrefix = {arXiv},
       eprint = {2505.20535},
 primaryClass = {stat.ML},
       adsurl = {https://ui.adsabs.harvard.edu/abs/2025arXiv250520535Z},
url={https://arxiv.org/html/2505.20535v1},
}

@article{Lopez-Fogliani:2024gzj,
doi = {10.1088/1475-7516/2025/01/057},
url = {https://dx.doi.org/10.1088/1475-7516/2025/01/057},
year = {2025},
month = {jan},
publisher = {IOP Publishing},
volume = {2025},
number = {01},
pages = {057},
author = {López-Fogliani, Daniel E. and Perez, Andres D. and de Austri, Roberto Ruiz},
title = {Insights into dark matter direct detection experiments: decision trees versus deep learning},
journal = {Journal of Cosmology and Astroparticle Physics}
}

@article{JMLR:v9:vandermaaten08a,
  author  = {Laurens van der Maaten and Geoffrey Hinton},
  title   = {Visualizing Data using t-SNE},
  journal = {Journal of Machine Learning Research},
  year    = {2008},
  volume  = {9},
  number  = {86},
  pages   = {2579--2605},
  url     = {http://jmlr.org/papers/v9/vandermaaten08a.html}
}

@article{DARWIN:2023uje,
   title={Cosmogenic background simulations for neutrinoless double beta decay with the DARWIN observatory at various underground sites},
   volume={84},
   ISSN={1434-6052},
   url={http://dx.doi.org/10.1140/epjc/s10052-023-12298-w},
   DOI={10.1140/epjc/s10052-023-12298-w},
   number={1},
   journal={The European Physical Journal C},
   publisher={Springer Science and Business Media LLC},
   author={DARWIN-Collaboration},
   year={2024},
   month=jan }

@article{Strigari_2009,
doi = {10.1088/1367-2630/11/10/105011},
url = {https://dx.doi.org/10.1088/1367-2630/11/10/105011},
year = {2009},
month = {oct},
publisher = {},
volume = {11},
number = {10},
pages = {105011},
author = {Louis E Strigari},
title = {Neutrino coherent scattering rates at direct dark matter detectors},
journal = {New Journal of Physics}
}

@article{Schumann_2015,
	doi = {10.1088/1475-7516/2015/10/016},
  
	url = {https://doi.org/10.1088%2F1475-7516%2F2015%2F10%2F016},
  
	year = 2015,
	month = {oct},
  
	publisher = {{IOP} Publishing},
  
	volume = {2015},
  
	number = {10},
  
	pages = {016--016},
  
	author = {Marc Schumann and Laura Baudis and Lukas Bütikofer and Alexander Kish and Marco Selvi},
  
	title = {Dark matter sensitivity of multi-ton liquid xenon detectors},
  
	journal = {Journal of Cosmology and Astroparticle Physics}
}

@article{GEANT4:2002zbu,
    author = "Agostinelli, S. and others",
    collaboration = "GEANT4",
    title = "{GEANT4--a simulation toolkit}",
    reportNumber = "SLAC-PUB-9350, FERMILAB-PUB-03-339, CERN-IT-2002-003",
    doi = "10.1016/S0168-9002(03)01368-8",
url="https://www.sciencedirect.com/science/article/abs/pii/S0168900203013688",
    journal = "Nucl. Instrum. Meth. A",
    volume = "506",
    pages = "250--303",
    year = "2003"
}

@article{Abbasi_2022,
	doi = {10.1088/1748-0221/17/06/p06026},
        title={Framework and tools for the simulation and analysis of the radio emission from air showers at IceCube},
	url = {https://doi.org/10.1088%2F1748-0221%2F17%2F06%2Fp06026},
  
	year = 2022,
	month = {jun},
  
	publisher = {{IOP} Publishing},
  
	volume = {17},
  
	number = {06},
  
	pages = {P06026},
  
	author = {IceCube collaboration},
  
	title = {Framework and tools for the simulation and analysis of the radio emission from air showers at {IceCube}},
  
	journal = {Journal of Instrumentation}
}

@article{Carleo_2019,
  author    = {Giuseppe Carleo and Ignacio Cirac and Kyle Cranmer and Laurent Daudet and Maria Schuld and Naftali Tishby and Leslie Vogt-Maranto and Lenka Zdeborová},
  title     = {Machine learning and the physical sciences},
  journal   = {Reviews of Modern Physics},
url = "https://journals.aps.org/rmp/abstract/10.1103/RevModPhys.91.045002",
  volume    = {91},
  number    = {4},
  year      = {2019},
  pages     = {045002},
  doi       = {10.1103/RevModPhys.91.045002},
}

@article{xenoncollaboration2024xenonntwimpsearchsignal,
    author = "Aprile, E. and others",
    collaboration = "XENON",
    title = "{XENONnT WIMP Search: Signal \& Background Modeling and Statistical Inference}",
    eprint = "2406.13638",
    archivePrefix = "arXiv",
    primaryClass = "physics.data-an",
    month = "6",
    year = "2024",
doi = {10.48550/arXiv.2406.13638},
url="https://hal.science/hal-04659687",
}

@article{Aalbers_2023LZ,
   title={First Dark Matter Search Results from the LUX-ZEPLIN (LZ) Experiment},
   volume={131},
   ISSN={1079-7114},
   url={http://dx.doi.org/10.1103/PhysRevLett.131.041002},
   DOI={10.1103/physrevlett.131.041002},
   number={4},
   journal={Physical Review Letters},
   publisher={American Physical Society (APS)},
   author={LUX-ZEPLIN-collaboration},
   year={2023},
   month=jul }

@article{xlzdDB,
    author = "Aalbers, J. and others",
    collaboration = "XLZD",
    title = {The XLZD Design Book: Towards the Next-Generation Liquid Xenon Observatory for Dark Matter and Neutrino Physics},
    eprint = "2410.17137",
    archivePrefix = "arXiv",
    primaryClass = "hep-ex",
    url = "https://inspirehep.net/literature/2841888",
    month = "10",
    year = "2024"
}

@article{Peres_2023,
   title={SiPM array of Xenoscope, a full-scale DARWIN vertical demonstrator},
   volume={18},
   ISSN={1748-0221},
   url={http://dx.doi.org/10.1088/1748-0221/18/03/C03027},
   DOI={10.1088/1748-0221/18/03/c03027},
   number={03},
   journal={Journal of Instrumentation},
   publisher={IOP Publishing},
   author={Peres, R.},
   year={2023},
   month=mar, pages={C03027} }

@article{chepel2013liquid,
  title={Liquid noble gas detectors for low energy particle physics},
  author={Chepel, Vitaly and Ara{\'u}jo, Henrique},
  journal={Journal of Instrumentation},
  volume={8},
  number={04},
  pages={R04001},
  year={2013},
  publisher={IOP Publishing},
url={https://iopscience.iop.org/article/10.1088/1748-0221/8/04/R04001/pdf}
}

@article{ParticleDataGroup:2022pth,
    author = "Workman, R. L. and others",
    collaboration = "Particle Data Group",
    title = "{Review of Particle Physics}",
    doi = "10.1093/ptep/ptac097",
    journal = "PTEP",
    volume = "2022",
    pages = "083C01",
    year = "2022",
url={https://academic.oup.com/ptep/article/2022/8/083C01/6651666}
}

@INPROCEEDINGS{sikka2019closer,
  author={Sikka, Harshvardhan and Zhong, Weishun and Yin, Jun and Pehlevant, Cengiz},
  booktitle={2019 53rd Asilomar Conference on Signals, Systems, and Computers}, 
  title={A Closer Look at Disentangling in β-VAE}, 
  year={2019},
  volume={},
    url="https://ieeexplore.ieee.org/abstract/document/9048921?casa_token=beCDJ95S9SsAAAAA:yG8VCLJ40_QZ_Oa7XVQpW3Te4G2xHBKiUKskBYqpQNlPf1PlrvoQmKYEaPBL3DcakoyGeAFW9Q",
  number={},
  pages={888-895},
  doi={10.1109/IEEECONF44664.2019.9048921}}

@book{Goodfellow-et-al-2016,
    title={Deep Learning},
    author={Ian Goodfellow and Yoshua Bengio and Aaron Courville},
    publisher={MIT Press},
    url={http://www.deeplearningbook.org},
    year={2016}
}

@misc{kingma2022autoencoding,
      title={Auto-Encoding Variational Bayes}, 
      author={Diederik P Kingma and Max Welling},
      year={2022},
      eprint={1312.6114},
      archivePrefix={arXiv},
      primaryClass={stat.ML},
  publisher={Banff, Canada},
url="http://web2.cs.columbia.edu/~blei/fogm/2018F/materials/KingmaWelling2013.pdf"
}

@article{arxiv_sim_inf,
author = {Kumar, Ankit and Singh, Abhishek and Doshi, Kavi and Goyal, Anirudh},
title = {Compositional Score Modeling for Simulation-based Inference},
journal = {arXiv preprint arXiv:2209.14249},
year = {2022}
}

@article{Goodman:1984dc,
	author = {Goodman, Mark W. and Witten, Edward},
	doi = {10.1103/PhysRevD.31.3059},
	editor = {Srednicki, M. A.},
	journal = {Phys. Rev. D},
	pages = {3059},
	reportnumber = {Print-85-0030 (PRINCETON)},
	title = {{Detectability of Certain Dark Matter Candidates}},
	volume = {31},
url="https://journals.aps.org/prd/abstract/10.1103/PhysRevD.31.3059",
	year = {1985},
	url = {https://doi.org/10.1103/PhysRevD.31.3059}}

@article{Schmidhuber_2015,
	author = {Schmidhuber, J{\"u}rgen},
	doi = {10.1016/j.neunet.2014.09.003},
	issn = {0893-6080},
	journal = {Neural Networks},
	month = {Jan},
	pages = {85--117},
	publisher = {Elsevier BV},
	title = {Deep learning in neural networks: An overview},
	url = {http://dx.doi.org/10.1016/j.neunet.2014.09.003},
	volume = {61},
	year = {2015},
	url = {http://dx.doi.org/10.1016/j.neunet.2014.09.003}}

@misc{bank2021autoencoders,
	archiveprefix = {arXiv},
	author = {Dor Bank and Noam Koenigstein and Raja Giryes},
	eprint = {2003.05991},
	primaryclass = {cs.LG},
	title = {Autoencoders},
	year = {2021},
url="https://link.springer.com/chapter/10.1007/978-3-031-24628-9_16"}

@article{us_sup,
doi = {10.1088/1475-7516/2022/02/039},
url = {https://dx.doi.org/10.1088/1475-7516/2022/02/039},
year = {2022},
month = {feb},
publisher = {IOP Publishing},
volume = {2022},
number = {02},
pages = {039},
author = {Herrero-Garcia, Juan and Patrick, Riley and Scaffidi, Andre},
title = {A semi-supervised approach to dark matter
	searches in direct detection data with machine learning},
journal = {Journal of Cosmology and Astroparticle Physics}
}

@article{Coarasa:2022zak,
    author = "Coarasa, I. and others",
    title = "{Improving ANAIS-112 sensitivity to DAMA/LIBRA signal with machine learning techniques}",
    eprint = "2209.14113",
    archivePrefix = "arXiv",
    primaryClass = "astro-ph.IM",
    doi = "10.1088/1475-7516/2022/11/048",
    journal = "JCAP",
    volume = "11",
    pages = "048",
    year = "2022",
    note = "[Erratum: JCAP 06, E01 (2023)]",
url="https://iopscience.iop.org/article/10.1088/1475-7516/2022/11/048"
}

@article{Todarello:2023qrr,
    author = "Todarello, Elisa and Scaffidi, Andre and Regis, Marco and Taoso, Marco",
    title = "{Constraining below-threshold radio source counts with machine learning}",
    eprint = "2306.15720",
    archivePrefix = "arXiv",
    primaryClass = "astro-ph.IM",
    doi = "10.1088/1475-7516/2024/01/062",
    journal = "JCAP",
    volume = "01",
    pages = "062",
    year = "2024",
url="https://iopscience.iop.org/article/10.1088/1475-7516/2024/01/062"
}

@misc{builtjes2024attention,
      title={Attention to the strengths of physical interactions: Transformer and graph-based event classification for particle physics experiments}, 
      author={Luc Builtjes and Sascha Caron and Polina Moskvitina and Clara Nellist and Roberto Ruiz de Austri and Rob Verheyen and Zhongyi Zhang},
      year={2024},
      eprint={2211.05143},
      archivePrefix={arXiv},
      primaryClass={hep-ph},
url="https://inspirehep.net/literature/2180384"
}

@article{Cowan:2010js,
    author = "Cowan, Glen and Cranmer, Kyle and Gross, Eilam and Vitells, Ofer",
    title = "{Asymptotic formulae for likelihood-based tests of new physics}",
    eprint = "1007.1727",
    archivePrefix = "arXiv",
    primaryClass = "physics.data-an",
    doi = "10.1140/epjc/s10052-011-1554-0",
    journal = "Eur. Phys. J. C",
    volume = "71",
    pages = "1554",
    year = "2011",
    note = "[Erratum: Eur.Phys.J.C 73, 2501 (2013)]",
url="https://link.springer.com/article/10.1140/epjc/s10052-011-1554-0"
}

@article{Dorta2018Training,
  title={Training VAEs Under Structured Residuals},
  author={Garoe Dorta and Sara Vicente and Lourdes de Agapito and Neill D. F. Campbell and Ivor J. A. Simpson},
  year={2018},
  url={https://api.semanticscholar.org/CorpusID:4560603}
}

@article{XENON:2022vye,
    author = "Aprile, E. and others",
    collaboration = "XENON",
    title = "{The triggerless data acquisition system of the XENONnT experiment}",
    eprint = "2212.11032",
    archivePrefix = "arXiv",
    primaryClass = "physics.ins-det",
    doi = "10.1088/1748-0221/18/07/P07054",
    journal = "JINST",
    volume = "18",
    number = "07",
    pages = "P07054",
    year = "2023",
url="https://iopscience.iop.org/article/10.1088/1748-0221/18/07/P07054"
}

@misc{nest2018,
  title        = {NEST Version v2.3.12},
  year         = 2018,
  doi          = {10.5281/zenodo.1314499},
  url          = {https://doi.org/10.5281/zenodo.1314499}
}

@misc{tensorflow2015-whitepaper,
  author={Mart\'{\i}n Abadi and others},
  year={2015},
  eprint={1603.04467},
  archivePrefix={arXiv},
  primaryClass={cs.LG},
  url={https://www.tensorflow.org/}
}

@inproceedings{hewitt2024deep,
  title={Deep learning approaches for fast event reconstruction in the SNO+ scintillator phase and beyond},
  author={Hewitt, Cal and Anderson, Mark},
  booktitle={Neutrino Physics and Machine Learning 2024},
  year={2024},
  organization={ETH Zurich},
  url={https://indico.phys.ethz.ch/event/113/contributions/827/}
}

@article{Jiang:2024wph,
      title={Machine-Learning based photon counting for PMT waveforms and its application to the improvement of the energy resolution in large liquid scintillator detectors}, 
      author={Wei Jiang and Guihong Huang and Zhen Liu and Wuming Luo and Liangjian Wen and Jianyi Luo},
      year={2024},
      eprint={2405.18720},
      archivePrefix={arXiv},
      primaryClass={physics.ins-det},
      url={https://link.springer.com/article/10.1140/epjc/s10052-024-13724-3}, 
}

@article{Farrell:2024aah,
    author = "Farrell, Sophia and Bergevin, Marc and Bernstein, Adam",
    title = "{Physics-informed machine learning approaches to reactor antineutrino detection}",
    eprint = "2407.06139",
    archivePrefix = "arXiv",
    primaryClass = "physics.ins-det",
    reportNumber = "LLNL-JRNL-865846",
    month = "7",
    year = "2024",
url="https://ui.adsabs.harvard.edu/abs/2024arXiv240706139F/abstract"
}

@inproceedings{Kessler:2014jya,
    author = "Kessler, Gaudenz",
    collaboration = "XENON100",
    title = "{XENON100 and XENON1T Dark Matter Search with Liquid Xenon}",
    booktitle = "{20th International Conference on Particles and Nuclei}",
    doi = "10.3204/DESY-PROC-2014-04/109",
    pages = "357--360",
url="http://dx.doi.org/10.3204/DESY-PROC-2014-04/109",
    month = "9",
    year = "2014"
}

@phdthesis{weber2013,
title={Gentle Neutron Signals and Noble Background in the XENON100 Dark Matter Search Experiment},
author={Weber, Marcus},
year={2013},
school={Ruprecht-Karls-Universit{\"a}t Heidelberg},
url={https://core.ac.uk/download/pdf/161443046.pdf}
}

@inproceedings{vetter2024,
title={Probabilistic Position Reconstruction in the XENONnT Experiment},
author={Vetter, S},
year={2024},
booktitle={Neutrino Physics and Machine Learning (NPML)},
organization={ETH Zurich},
url={https://indico.phys.ethz.ch/event/113/contributions/890/}
}

@article{pancake,
doi = {10.1088/1748-0221/19/05/P05018},
url = {https://dx.doi.org/10.1088/1748-0221/19/05/P05018},
year = {2024},
month = {may},
publisher = {IOP Publishing},
volume = {19},
number = {05},
pages = {P05018},
author = {Brown, Adam and Fischer, Horst and Glade-Beucke, Robin and Grigat, Jaron and Kuger, Fabian and Lindemann, Sebastian and Luce, Tiffany and Masson, Darryl and Müller, Julia and Reininghaus, Jens and Schumann, Marc and Stevens, Andrew and Tönnies, Florian and Toschi, Francesco},
title = {PANCAKE: a large-diameter cryogenic test platform with a flat floor for next generation multi-tonne liquid xenon detectors},
journal = {Journal of Instrumentation}
}

@article{Akerib2022,
  title={Fast and flexible analysis of direct dark matter search data with machine learning},
  author={Akerib, D. S. and others (LUX Collaboration)},
  journal={Physical Review D},
  volume={106},
  number={7},
  pages={072009},
  year={2022},
  publisher={American Physical Society},
  url={https://journals.aps.org/prd/abstract/10.1103/PhysRevD.106.072009}
}

@article{XENONCollaboration:2023dar,
    author = "Aprile, E. and others",
    collaboration = "(XENON Collaboration)\textdagger{}\textdagger{}, XENON",
    title = "{Detector signal characterization with a Bayesian network in XENONnT}",
    eprint = "2304.05428",
    archivePrefix = "arXiv",
    primaryClass = "hep-ex",
    doi = "10.1103/PhysRevD.108.012016",
    journal = "Phys. Rev. D",
    volume = "108",
    number = "1",
    pages = "012016",
    year = "2023",
url="https://journals.aps.org/prd/abstract/10.1103/PhysRevD.108.012016"
}

@article{gelman2013bayesian,
  title={Bayesian Data Analysis},
  author={Gelman, Andrew and Carlin, John B and Stern, Hal S and Dunson, David B and Vehtari, Aki and Rubin, Donald B},
  year={2013},
  publisher={Chapman and Hall/CRC},
url="https://sites.stat.columbia.edu/gelman/book/BDA3.pdf"
}

@article{DarkSide-50:2023fcw,
    author = "Agnes, P. and others",
    collaboration = "DarkSide-50",
    title = "{Search for low mass dark matter in DarkSide-50: the bayesian network approach}",
    eprint = "2302.01830",
    archivePrefix = "arXiv",
    primaryClass = "hep-ex",
    reportNumber = "FERMILAB-PUB-23-043-AD-CSAID-ND",
    doi = "10.1140/epjc/s10052-023-11410-4",
    journal = "Eur. Phys. J. C",
    volume = "83",
    pages = "322",
    year = "2023",
url="https://link.springer.com/article/10.1140/epjc/s10052-023-11410-4"
}

@article{OHare:2016pjy,
    author = "O'Hare, Ciaran A. J.",
    title = "{Dark matter astrophysical uncertainties and the neutrino floor}",
    eprint = "1604.03858",
    archivePrefix = "arXiv",
    primaryClass = "astro-ph.CO",
    doi = "10.1103/PhysRevD.94.063527",
    journal = "Phys. Rev. D",
    volume = "94",
    number = "6",
    pages = "063527",
    year = "2016",
url="https://journals.aps.org/prd/abstract/10.1103/PhysRevD.94.063527"
}

@article{Blance_2019,
   title={Adversarially-trained autoencoders for robust unsupervised new physics searches},
   volume={2019},
   ISSN={1029-8479},
   url={http://dx.doi.org/10.1007/JHEP10(2019)047},
   DOI={10.1007/jhep10(2019)047},
   number={10},
   journal={Journal of High Energy Physics},
   publisher={Springer Science and Business Media LLC},
   author={Blance, Andrew and Spannowsky, Michael and Waite, Philip},
   year={2019},
   month=oct }

@article{Blance_2021,
   title={Quantum machine learning for particle physics using a variational quantum classifier},
   volume={2021},
   ISSN={1029-8479},
   url={http://dx.doi.org/10.1007/JHEP02(2021)212},
   DOI={10.1007/jhep02(2021)212},
   number={2},
   journal={Journal of High Energy Physics},
   publisher={Springer Science and Business Media LLC},
   author={Blance, Andrew and Spannowsky, Michael},
   year={2021},
   month=feb }

@article{gong2018causal,
  title={Causal Generative Domain Adaptation Networks},
  author={Mingming Gong and Kun Zhang and Biwei Huang and Clark Glymour and Dacheng Tao and K. Batmanghelich},
  journal={ArXiv},
  year={2018},
  volume={abs/1804.04333},
  url={https://api.semanticscholar.org/CorpusID:4807438}
}

@article{akerib2016lowenergy,
    author = "Akerib, D. S. and others",
    collaboration = "LUX",
    title = "{Low-energy (0.7-74 keV) nuclear recoil calibration of the LUX dark matter experiment using D-D neutron scattering kinematics}",
    eprint = "1608.05381",
    archivePrefix = "arXiv",
    primaryClass = "physics.ins-det",
    month = "8",
    year = "2016",
url={https://www.researchgate.net/publication/306285462_Low-energy_07-74_keV_nuclear_recoil_calibration_of_the_LUX_dark_matter_experiment_using_D-D_neutron_scattering_kinematics}
}

@article{lenardo2019ionization,
    author = "Lenardo, Brian and others",
    title = "{Measurement of the ionization yield from nuclear recoils in liquid xenon between 0.3 - 6 keV with single-ionization-electron sensitivity}",
    eprint = "1908.00518",
    archivePrefix = "arXiv",
    primaryClass = "physics.ins-det",
    month = "8",
    year = "2019",
url = {https://www.researchgate.net/publication/301572459_Improved_Limits_on_Scattering_of_Weakly_Interacting_Massive_Particles_from_Reanalysis_of_2013_LUX_Data},
}

@misc{hermans2021averting,
      title={A Trust Crisis In Simulation-Based Inference? Your Posterior Approximations Can Be Unfaithful}, 
      author={Joeri Hermans and Arnaud Delaunoy and François Rozet and Antoine Wehenkel and Volodimir Begy and Gilles Louppe},
      year={2022},
      eprint={2110.06581},
      archivePrefix={arXiv},
      primaryClass={stat.ML},
      url={https://arxiv.org/abs/2110.06581}, 
}

@article{edwards1985exact,
author = {Don Edwards and},
title = {Exact simulation-based inference: A survey, with additions},
journal = {Journal of Statistical Computation and Simulation},
volume = {22},
number = {3-4},
pages = {307--326},
year = {1985},
publisher = {Taylor \& Francis},
doi = {10.1080/00949658508810853},


URL = { 
    
        https://doi.org/10.1080/00949658508810853
    
    

},
eprint = { 
    
        https://doi.org/10.1080/00949658508810853
    
    

}

}

@article{Aalbers_2023,
doi = {10.1088/1361-6471/ac841a},
title={A next-generation liquid xenon observatory for dark matter and neutrino physics},
url = {https://dx.doi.org/10.1088/1361-6471/ac841a},
year = {2022},
month = {dec},
publisher = {IOP Publishing},
volume = {50},
number = {1},
pages = {013001},
author = {Aalbers, J and others},
title = {A next-generation liquid xenon observatory for dark matter and neutrino physics},
journal = {Journal of Physics G: Nuclear and Particle Physics}
}

@article{Aalbers:2016jon,
   title={DARWIN: towards the ultimate dark matter detector},
   volume={2016},
   ISSN={1475-7516},
   url={http://dx.doi.org/10.1088/1475-7516/2016/11/017},
   DOI={10.1088/1475-7516/2016/11/017},
   number={11},
   journal={Journal of Cosmology and Astroparticle Physics},
   publisher={IOP Publishing},
   author={Aalbers, J. and others},
   year={2016},
   month=nov, pages={017–017} }

@article{Lai:2023qub,
    author = "Lai, Michela",
    collaboration = "DEAP-3600",
    title = "{Recent results from DEAP-3600}",
    eprint = "2302.14484",
    archivePrefix = "arXiv",
    primaryClass = "hep-ex",
    doi = "10.1088/1748-0221/18/02/C02046",
url="https://iopscience.iop.org/article/10.1088/1748-0221/18/02/C02046",
    journal = "JINST",
    volume = "18",
    number = "02",
    pages = "C02046",
    year = "2023"
}

@article{XENON:2023cxc,
    author = "Aprile, E. and others",
    collaboration = "XENON",
    title = "{First Dark Matter Search with Nuclear Recoils from the XENONnT Experiment}",
    eprint = "2303.14729",
url="https://journals.aps.org/prl/abstract/10.1103/PhysRevLett.131.041003",
    archivePrefix = "arXiv",
    primaryClass = "hep-ex",
    doi = "10.1103/PhysRevLett.131.041003",
    journal = "Phys. Rev. Lett.",
    volume = "131",
    number = "4",
    pages = "041003",
    year = "2023"
}

@article{PandaX-4T:2021bab,
    author = "Meng, Yue and others",
    collaboration = "PandaX-4T",
    title = "{Dark Matter Search Results from the PandaX-4T Commissioning Run}",
    eprint = "2107.13438",
    archivePrefix = "arXiv",
    primaryClass = "hep-ex",
    doi = "10.1103/PhysRevLett.127.261802",
    url="https://journals.aps.org/prl/abstract/10.1103/PhysRevLett.127.261802",
    journal = "Phys. Rev. Lett.",
    volume = "127",
    number = "26",
    pages = "261802",
    year = "2021"
}

@article{Calvo:2016hve,
	archiveprefix = {arXiv},
	author = {Calvo, J. and others},
	collaboration = {ArDM},
url="https://iopscience.iop.org/article/10.1088/1475-7516/2017/03/003",
	doi = {10.1088/1475-7516/2017/03/003},
	eprint = {1612.06375},
	journal = {JCAP},
	pages = {003},
	primaryclass = {physics.ins-det},
	title = {{Commissioning of the ArDM experiment at the Canfranc underground laboratory: first steps towards a tonne-scale liquid argon time projection chamber for Dark Matter searches}},
	volume = {03},
	year = {2017},}

@article{Aalseth:2017fik,
	archiveprefix = {arXiv},
	author = {Aalseth, C.E. and others},
	doi = {10.1140/epjp/i2018-11973-4},
	eprint = {1707.08145},
    url = "https://link.springer.com/article/10.1140/epjp/i2018-11973-4",
	journal = {Eur. Phys. J. Plus},
	pages = {131},
	primaryclass = {physics.ins-det},
	reportnumber = {FERMILAB-PUB-17-298-PPD},
	title = {{DarkSide-20k: A 20 tonne two-phase LAr TPC for direct dark matter detection at LNGS}},
	volume = {133},
	year = {2018},}

@article{Zhang:2019ryt,
	archiveprefix = {arXiv},
	author = {Zhang, Xinyue and Wang, Yanfang and Zhang, Wei and Sun, Yueqiu and He, Siyu and Contardo, Gabriella and Villaescusa-Navarro, Francisco and Ho, Shirley},
	eprint = {1902.05965},
	month = {2},
	primaryclass = {astro-ph.CO},
url="https://ui.adsabs.harvard.edu/abs/2019arXiv190205965Z/abstract",
	title = {{From Dark Matter to Galaxies with Convolutional Networks}},
	year = {2019}}

@Article{Khosa:2020qrz,
	title={{Anomaly Awareness}},
	author={Charanjit K. Khosa and Veronica Sanz},
	journal={SciPost Phys.},
	volume={15},
	pages={053},
	year={2023},
	publisher={SciPost},
	doi={10.21468/SciPostPhys.15.2.053},
	url={https://scipost.org/10.21468/SciPostPhys.15.2.053},
}

@article{Lucie-Smith:2019hdl,
	archiveprefix = {arXiv},
	author = {Lucie-Smith, Luisa and Peiris, Hiranya V. and Pontzen, Andrew},
	doi = {10.1093/mnras/stz2599},
	eprint = {1906.06339},
	journal = {Mon. Not. Roy. Astron. Soc.},
	number = {1},
	pages = {331--342},
	primaryclass = {astro-ph.CO},
	title = {{An interpretable machine learning framework for dark matter halo formation}},
	volume = {490},
	year = {2019},
	url = {https://doi.org/10.1093/mnras/stz2599}}

@article{Bernardini:2019bmd,
	archiveprefix = {arXiv},
	author = {Bernardini, Mauro and Mayer, Lucio and Reed, Darren and Feldmann, Robert},
	doi = {10.1093/mnras/staa1911},
	eprint = {1912.04299},
	journal = {Mon. Not. Roy. Astron. Soc.},
	number = {4},
	pages = {5116--5125},
	primaryclass = {astro-ph.CO},
	title = {{Predicting dark matter halo formation in N-body simulations with deep regression networks}},
	volume = {496},
	year = {2020},
	url = {https://doi.org/10.1093/mnras/staa1911}}

@article{Farina:2018fyg,
  title = {Searching for new physics with deep autoencoders},
  author = {Farina, Marco and Nakai, Yuichiro and Shih, David},
  journal = {Phys. Rev. D},
  volume = {101},
  issue = {7},
  pages = {075021},
  numpages = {13},
  year = {2020},
  month = {Apr},
  publisher = {American Physical Society},
  doi = {10.1103/PhysRevD.101.075021},
  url = {https://link.aps.org/doi/10.1103/PhysRevD.101.075021}
}

@article{Heimel:2018mkt,
	archiveprefix = {arXiv},
	author = {Heimel, Theo and Kasieczka, Gregor and Plehn, Tilman and Thompson, Jennifer M.},
	doi = {10.21468/SciPostPhys.6.3.030},
	eprint = {1808.08979},
	journal = {SciPost Phys.},
	number = {3},
	pages = {030},
	primaryclass = {hep-ph},
	slaccitation = {%%CITATION = ARXIV:1808.08979;%%},
	title = {{QCD or What?}},
	volume = {6},
	year = {2019},
	url = {https://doi.org/10.21468/SciPostPhys.6.3.030}}

@article{Kuusela:2011aa,
	archiveprefix = {arXiv},
	author = {Kuusela, Mikael and Vatanen, Tommi and Malmi, Eric and Raiko, Tapani and Aaltonen, Timo and Nagai, Yoshikazu},
	booktitle = {{Proceedings, 14th International Workshop on Advanced Computing and Analysis Techniques in Physics Research (ACAT 2011): Uxbridge, UK, September 5-9, 2011}},
	doi = {10.1088/1742-6596/368/1/012032},
	eprint = {1112.3329},
	journal = {J. Phys. Conf. Ser.},
	pages = {012032},
	primaryclass = {physics.data-an},
	slaccitation = {%%CITATION = ARXIV:1112.3329;%%},
	title = {{Semi-Supervised Anomaly Detection - Towards Model-Independent Searches of New Physics}},
	volume = {368},
	year = {2012},
	url = {https://doi.org/10.1088/1742-6596/368/1/012032}}

@article{cerri2019variational,
	author = {Cerri, Olmo and Nguyen, Thong Q and Pierini, Maurizio and Spiropulu, Maria and Vlimant, Jean-Roch},
	journal = {Journal of High Energy Physics},
	number = {5},
	pages = {36},
	publisher = {Springer},
	title = {Variational autoencoders for new physics mining at the large hadron collider},
	volume = {2019},
url="https://link.springer.com/article/10.1007/JHEP05(2019)036",
	year = {2019}}

@article{knapp2020adversarially,
	author = {Knapp, Oliver and Dissertori, Guenther and Cerri, Olmo and Nguyen, Thong Q and Vlimant, Jean-Roch and Pierini, Maurizio},
	journal = {arXiv preprint arXiv:2005.01598},
url="https://link.springer.com/article/10.1140/epjp/s13360-021-01109-4",
	title = {Adversarially Learned Anomaly Detection on CMS Open Data: re-discovering the top quark},
	year = {2020}}

@article{Andreassen:2020nkr,
	archiveprefix = {arXiv},
	author = {Andreassen, Anders and Nachman, Benjamin and Shih, David},
	doi = {10.1103/PhysRevD.101.095004},
	eprint = {2001.05001},
	journal = {Phys. Rev. D},
	number = {9},
	pages = {095004},
	primaryclass = {hep-ph},
	title = {{Simulation Assisted Likelihood-free Anomaly Detection}},
	volume = {101},
	year = {2020},
	url = {https://doi.org/10.1103/PhysRevD.101.095004}}

@inproceedings{Arthurs2024,
  title={Application of machine learning for anomaly detection and background discrimination in LZ data},
  author={Maris Arthurs},
  booktitle={Conference on Science at the Sanford Underground Research Facility},
  year={2024},
  address={SD Mines, South Dakota, USA},
  month={May},
  url={https://indico.sanfordlab.org/event/68/contributions/1323/}
}

@article{Nachman:2020lpy,
	archiveprefix = {arXiv},
	author = {Nachman, Benjamin and Shih, David},
	doi = {10.1103/PhysRevD.101.075042},
	eprint = {2001.04990},
	journal = {Phys. Rev. D},
	pages = {075042},
	primaryclass = {hep-ph},
	title = {{Anomaly Detection with Density Estimation}},
	volume = {101},
	year = {2020},
	url = {https://doi.org/10.1103/PhysRevD.101.075042}}

@article{Collins:2019jip,
	archiveprefix = {arXiv},
	author = {Collins, Jack H. and Howe, Kiel and Nachman, Benjamin},
	doi = {10.1103/PhysRevD.99.014038},
	eprint = {1902.02634},
	journal = {Phys. Rev. D},
	number = {1},
	pages = {014038},
	primaryclass = {hep-ph},
	reportnumber = {FERMILAB-PUB-18-733-T},
	title = {{Extending the search for new resonances with machine learning}},
	volume = {99},
	year = {2019},
	url = {https://doi.org/10.1103/PhysRevD.99.014038}}

@article{Dery:2018dqr,
	author = {Dery, Lucio Mwinmaarong and Nachman, Benjamin and Rubbo, Francesco and Schwartzman, Ariel},
	doi = {10.1088/1742-6596/1085/4/042006},
	journal = {J. Phys. Conf. Ser.},
	number = {4},
	pages = {042006},
	title = {{Weakly Supervised Classification For High Energy Physics}},
	volume = {1085},
	year = {2018},
	url = {https://doi.org/10.1088/1742-6596/1085/4/042006}}

@article{Collins:2018epr,
	archiveprefix = {arXiv},
	author = {Collins, Jack H. and Howe, Kiel and Nachman, Benjamin},
	doi = {10.1103/PhysRevLett.121.241803},
	eprint = {1805.02664},
	journal = {Phys. Rev. Lett.},
	number = {24},
	pages = {241803},
	primaryclass = {hep-ph},
	reportnumber = {FERMILAB-PUB-18-180-T},
	title = {{Anomaly Detection for Resonant New Physics with Machine Learning}},
	volume = {121},
	year = {2018},
	url = {https://doi.org/10.1103/PhysRevLett.121.241803}}

@article{Otten:2019hhl,
    author = "Otten, Sydney and Caron, Sascha and de Swart, Wieske and van Beekveld, Melissa and Hendriks, Luc and van Leeuwen, Caspar and Podareanu, Damian and Ruiz de Austri, Roberto and Verheyen, Rob",
    title = "{Event Generation and Statistical Sampling for Physics with Deep Generative Models and a Density Information Buffer}",
    eprint = "1901.00875",
    archivePrefix = "arXiv",
    primaryClass = "hep-ph",
    doi = "10.1038/s41467-021-22616-z",
    journal = "Nature Commun.",
    volume = "12",
    number = "1",
    pages = "2985",
    year = "2021",
url="https://www.nature.com/articles/s41467-021-22616-z",
}

@article{vanBeekveld:2020txa,
    author = "van Beekveld, Melissa and Caron, Sascha and Hendriks, Luc and Jackson, Paul and Leinweber, Adam and Otten, Sydney and Patrick, Riley and Ruiz De Austri, Roberto and Santoni, Marco and White, Martin",
    title = "{Combining outlier analysis algorithms to identify new physics at the LHC}",
    eprint = "2010.07940",
    archivePrefix = "arXiv",
    primaryClass = "hep-ph",
    doi = "10.1007/JHEP09(2021)024",
    journal = "JHEP",
    volume = "09",
    pages = "024",
    year = "2021",
url="https://link.springer.com/article/10.1007/JHEP09(2021)024"
}
